\documentclass[sigconf, nonacm, pdfa]{acmart}

\newcommand\vldbdoi{XX.XX/XXX.XX}
\newcommand\vldbpages{XXX - XXX}
\newcommand\vldbvolume{X}
\newcommand\vldbissue{X}
\newcommand\vldbyear{2023}

\newcommand\vldbauthors{\authors}
\newcommand\vldbtitle{\shorttitle} 
\newcommand\vldbavailabilityurl{https://github.com/ds2-lab/aca-dlis}
\newcommand\vldbpagestyle{plain}

\usepackage[a-2b]{pdfx}
\usepackage[normalem]{ulem}
\usepackage{comment}

\usepackage{url}
\usepackage{listings}
\usepackage{epstopdf}
\usepackage{color}
\usepackage{subfigure}
\usepackage{xspace}
\usepackage{latexsym}
\makeatletter
\newif\if@restonecol
\makeatother

\usepackage[ruled, shortend, linesnumbered, noline]{algorithm2e}
\usepackage{algorithmic}
\usepackage{enumerate}

\usepackage{paralist}

\usepackage{makecell}

\usepackage{fontawesome}

\usepackage{wrapfig}
\usepackage{setspace}
\usepackage{blindtext}
\usepackage{amsfonts}
\usepackage{multirow}
\usepackage{pifont}
\usepackage{mathrsfs}
\usepackage{mathtools}
\usepackage{relsize}
\usepackage{comment}
\usepackage{footnote}
\usepackage{tikz}
\usetikzlibrary{calc}
\makesavenoteenv{tabular}
\makesavenoteenv{table}
\usepackage{soul}

\usepackage{framed}  
\colorlet{shadecolor}{gray!20}

\usepackage{pifont}
\newcommand{\cmark}{\ding{51}}%

\usepackage{makecell}
\usepackage{multirow}
\usepackage{booktabs}
\usepackage{siunitx}
\usepackage{enumitem}

\usepackage{zref}
\usepackage{cleveref}
\crefformat{section}{\S#2#1#3}

\begin{document}
\title{Algorithmic Complexity Attacks on Dynamic Learned Indexes} 



\author{Rui Yang}
\affiliation{%
  \institution{{\it University of Virginia}}
}
\email{qgh4hm@virginia.edu}

\author{Evgenios M. Kornaropoulos}
\affiliation{%
  \institution{{\it George Mason University}}
}
\email{evgenios@gmu.edu}

\author{Yue Cheng}
\affiliation{%
  \institution{{\it University of Virginia}}
}
\email{mrz7dp@virginia.edu}


\settopmatter{printfolios=true}

\begin{abstract}
\label{sec:abstract}
 Learned Index Structures (LIS) view a sorted index as a model that learns the data distribution, takes a data element key as input, and outputs the predicted position of the key. The original LIS can only handle lookup operations with no support for updates, rendering it impractical to use for typical workloads. To address this limitation, 
recent studies have focused on designing efficient dynamic learned indexes. ALEX, as the first and one of the representative dynamic learned index structures, enables dynamism by incorporating a series of design choices, including adaptive key space partitioning, dynamic model retraining, and sophisticated engineering and policies that prioritize read/write performance. While these design choices offer improved average-case performance, the emphasis on flexibility and performance increases the attack surface by allowing adversarial behaviors that maximize ALEX's memory space and time complexity in worst-case scenarios. 

In this work, we present the first systematic investigation of algorithmic complexity attacks (ACAs) targeting the worst-case scenarios of ALEX.
We introduce new ACAs that fall into two categories, space ACAs
and time ACAs, which target the memory space and time complexity, respectively. 
First, our space ACA on data nodes exploits ALEX's gapped array layout and uses Multiple-Choice Knapsack (MCK) to generate an optimal adversarial insertion plan for maximizing the memory consumption at the data node level. 
Second, our space ACA on internal nodes exploits ALEX's catastrophic cost mitigation mechanism,
causing an out-of-memory (OOM) error with only a few hundred adversarial insertions. 
Third, our time ACA generates pathological insertions to increase the disparity between the actual key distribution and the linear models of data nodes, deteriorating the runtime performance by up to $1,641\times$ compared to ALEX operating under legitimate workloads.
    
\end{abstract}

\maketitle


\pagestyle{\vldbpagestyle}
\begingroup\small\noindent\raggedright\textbf{PVLDB Reference Format:}\\
\vldbauthors. \vldbtitle. PVLDB, \vldbvolume(\vldbissue): \vldbpages, \vldbyear.\\
\href{https://doi.org/\vldbdoi}{doi:\vldbdoi}
\endgroup
\begingroup
\renewcommand\thefootnote{}\footnote{\noindent
This work is licensed under the Creative Commons BY-NC-ND 4.0 International License. Visit \url{https://creativecommons.org/licenses/by-nc-nd/4.0/} to view a copy of this license. For any use beyond those covered by this license, obtain permission by emailing \href{mailto:info@vldb.org}{info@vldb.org}. Copyright is held by the owner/author(s). Publication rights licensed to the VLDB Endowment. \\
\raggedright Proceedings of the VLDB Endowment, Vol. \vldbvolume, No. \vldbissue\ %
ISSN 2150-8097. \\
\href{https://doi.org/\vldbdoi}{doi:\vldbdoi} \\
}\addtocounter{footnote}{-1}\endgroup

\ifdefempty{\vldbavailabilityurl}{}{
\vspace{.3cm}
\begingroup\small\noindent\raggedright\textbf{PVLDB Artifact Availability:}\\
The source code, data, and/or other artifacts have been made available at \url{\vldbavailabilityurl}.
\endgroup
}

\section{Introduction}
\label{sec:intro}

Index structures are essential to database systems for efficient data storage and retrieval. Traditional database index structures precisely organize each and every data item to balance tradeoffs in query performance and memory efficiency. Recent work by Kraska, Beutel, Chi, Dean, and Polyzotis~\cite{lis_sigmod18} proposes to replace the traditional database index with a hierarchy of machine learning (ML) models, which is called \emph{Learned Index Structure} (LIS). The insight is that a sorted index can be viewed as a model, which learns the data distribution, takes a data element key as input, and outputs the location of the key stored in a storage medium (e.g., a disk). Under LIS, a  query starts by consulting a root model at the top of the hierarchy in order to predict the child model to use, and performs a model traversal downwards the model hierarchy until it reaches a leaf node; the leaf node is also a model that will return the predicted position of the key from a densely packed array. 

The very first LIS proposed in the literature could only handle read-only data with no support for update operations. 
This limitation made LIS unusable for dynamic read-write workloads. 
To address this gap, various dynamic index structures have been introduced, including ALEX~\cite{alex_sigmod20}, PGM~\cite{pgm_vldb20}, LIPP~\cite{lipp_vldb21}, and APEX~\cite{apex_vldb21}, among others. 
In our work, we select ALEX~\cite{alex_sigmod20} as our target for our algorithmic complexity attacks (ACAs), because not only is ALEX the \emph{first} dynamic learned index (DLIS), but also
(1)~ALEX has \emph{qualitative features} that make it a strong candidate for real deployment,
and (2)~there is strong \emph{quantitative evidence} that ALEX is the \emph{most efficient and well-rounded} DLIS that can support real workloads, see the recent benchmark studies~\cite{lis_benchmarking_vldb22, lis_benchmarking_vldb23} that rank ALEX highly across all tested performance metrics. 
Similar to LIS, ALEX organizes models in a tree structure and serves requests by traversing the tree from top to bottom. Unlike LIS, ALEX allows both the internal nodes and leaf nodes (data nodes that store the data element) to dynamically grow and shrink at difference rates. To achieve efficient lookup and insertion operations, ALEX uses an array with gaps (called a \emph{gapped array}). ALEX also introduces heuristic-guided, dynamic node expansion/splitting mechanisms, paired with model retraining, to adapt to workload changes. 

The dynamism of ALEX over LIS is achieved by enabling (1)~adaptive key space repartitioning, (2)~dynamic model retraining, and (3)~sophisticated engineering that prioritizes performance. 
While these design choices offer improved average-case performance, the emphasis on flexibility and performance increases the attack surface by allowing adversarial behaviors that maximize ALEX's space and time complexity in \emph{worst-case scenarios}. 
These worst-case scenarios can be exploited by a type of attack method called algorithmic complexity attacks (ACAs)~\cite{aca_usenixsec03, aca_intro}. 
ACAs are a class of Denial-of-Service (DoS) attacks, where an attacker uses a small amount of adversarial inputs to induce a large amount of work in the target system, pushing the system into consuming all available resources.

The key observation of this paper is that  \emph{ALEX's various design choices to handle dynamic workload} (including the gapped array layout, the cost models to guide node resizing and model retraining) \emph{lack worst-case guarantees, which can be exploited by attackers.} 

\noindent\textbf{Our Contributions. } 
In this paper, we perform a security analysis on ALEX's design choices and present three new ACAs.
To the best of our knowledge, our work is the \emph{first to systematically analyze algorithmic complexity vulnerabilities} of dynamic learned indexes. 
The three ACAs fall into two categories,
which target the space and time complexity aspects of the ALEX index structure. 

\noindent$\bullet$ The first category of attacks that we introduce concerns ACA that targets the memory space complexity dimension of ALEX. 
We discover a set of adversarial input cases that cause significant memory consumption increases at both the data node level and internal node level. 
First, our space ACA on data nodes exploits the design that ALEX uses a gapped array layout with dispersed empty space called gaps for performance to perform a memory space complexity attack. 
This attack models ALEX's global data node structure as a knapsack problem and uses Multiple-Choice Knapsack (MCK) to generate an optimal insertion plan for maximizing the memory consumption at the data node level.
Second, our space ACA on internal nodes issues duplicate keys to trigger ALEX's catastrophic cost mitigation mechanism, where ALEX needs to recursively split a data node with the hope to bring the duplicate-key-induced cost down to below the threshold; the catastrophic outcome of this attack is that, with just a small amount of insertions for duplicate keys, ALEX repeatedly applies recursive-split operations and exponentially increases the size of the internal node's pointer array, which, in turn, exhausts all available memory resources on the host. 

\noindent$\bullet$ 
The second category of attacks is time ACAs, wherein
we design an attack method aimed at increasing the number of model-retrain operations.
This attack generates pathological insertions  that consist of consecutive keys to deviate the actual key distribution from the linear model of a set of randomly chosen (under black-box attacks) or deliberately chosen (under white-box or gray-box attacks) data nodes. The outcome of this attack is a dramatically increased runtime performance overhead caused by model retraining.

\noindent$\bullet$
Our ACAs expose vulnerabilities that contradict the intended benefits of space efficiency and performance, which are the original design goals of LIS~\cite{lis_sigmod18}, thereby stressing the need for new designs of dynamic learned indexes with worst-case guarantees.

\begin{table}[t]
\scriptsize
    \centering
    \caption{A summary of targeted ALEX designs and characteristics of the proposed ACAs.}
    \vspace{-20pt}  
    \resizebox{\columnwidth}{!}{
    \begin{tabular}{|c|c|c|c|c|}
        \cline{3-5}
        \multicolumn{1}{c}{} & {} &  \bfseries\makecell[c]{Data-Node \\ Space ACA} &  \bfseries\makecell[c]{Internal-Node \\ Space ACA} & {\bf Time ACA} \\
        \hline
        \hline
         {} & \makecell[c]{
         Memory\\ Over-provisioning
         } & {\cmark} & {} & {} \\ \cline{2-5}
         {\bf Exploited Design Choice} & \makecell[c]{
         ML Prediction\\ Error Mitigation
         } & {} & {\cmark} & {\cmark} \\  \cline{2-5}
         {} &  \makecell[c]{Keyspace \\ Partitioning} & {} & {\cmark} & {} \\
        \hline
        \hline
        \multirow{2}{*}{\bfseries\makecell[c]{Targeted Component}}  & {Internal Node} & {} & {\cmark} & {} \\ \cline{2-5}
            & {Data Node} & {\cmark} & {} & {\cmark} \\
           \hline
        \hline                                     
        {} & {Black-box} & {} & {\cmark} & {\cmark} \\ \cline{2-5}
        {\bf Attack Setting} & {Gray-box} & {\cmark} & {} & {\cmark} \\ \cline{2-5}
        {} & {White-box} & {\cmark} & {} & {\cmark} \\ \cline{2-5}
           \hline
        \hline    
        \multirow{2}{*}{\bf Targeted Resource} & {Memory (space)} & {\cmark} & {\cmark} &{} \\ \cline{2-5}
        {} & {CPU (time)} & {} & {} &{\cmark} \\ \cline{2-5} 
        \hline  
    \end{tabular}
    }
    \label{table:aca_fm}
    \vspace{-4.8pt}
\end{table}
\noindent\textbf{Summary of Our Proposed ACAs.} 
In this work, we propose three different ACAs and summarize their features in Table~\ref{table:aca_fm}.
For each ACA, we explain the exploited ALEX design choices, the targeted components, the attack settings, and the target resources. The details are as follows:
(1) Our space ACA on data nodes is designed to target the memory space resource of the target host. It exploits the memory over-provisioning design choice of ALEX to maximize ALEX's memory consumption by expanding the capacity of data nodes. The effectiveness of this ACA depends on the key distribution of the workload and the adversarial budget size. It can be mounted in both white-box and gray-box settings.
(2) Our space ACA on internal nodes also targets the memory space resource but with a different approach. This attack aims to exhaust the target host's memory resources, by causing an OOM event. It exploits two design components of ALEX: recursive splitting of a data node and key-space based partitioning that halves the keyspace regardless of how keys are distributed in it. This attack requires the target database to support duplicate keys and can be mounted in a black-box setting.
(3) Our time ACA targets the host's CPU resource, to maximize the number of retrains to deteriorate the performance of the index. The effectiveness of this attack can be affected by the workload's key distribution, the read/write ratio, and the adversarial budget size. It can be mounted in all three settings.

\noindent\textbf{Overview of Our Findings. }We conduct an extensive set of experiments on real-world datasets. For space ACA on data nodes, we increase the memory usage by up to $31\%$ compared to baseline ALEX with a budget size ranging from $1\%$ to $30\%$ of various workload sizes. 
In the case of space ACA on internal nodes, our internal-node ACA is capable of depleting the entire memory resources of the host machine with just a few hundred of adversarial insertions. 
Our time ACAs cause a throughput degradation of  ALEX by up to $1,641\times$ with a small adversarial budget (ranging from $2.5\%$ to $10\%$ of the workload size for write-heavy workloads and $0.5\%$ to $2\%$ of the workload size for read-heavy workloads). 
\vspace{-2pt}
\section{Preliminaries}
\label{sec:pre}

\subsection{Static Learned Index Structures}
Kraska et~al.~\cite{lis_sigmod18} propose to replace the traditional index structures with a hierarchy of learned models that they call \emph{Learned Index Structures} (LIS). 
The insight of LIS is that the task of locating a key $k$ within a set of linearly ordered keys can be reduced to approximating the probability that a randomly chosen key would take a value less or equal to $k$, i.e., $Pr(X\leq k) = Rank(k)/n$, where $X$ is the random variable that follows the empirical distribution of the $n$ keys and $Rank(.)$ is a function that takes a key as an input and outputs its relative position among the $n$ keys. 
This probability is captured by the cumulative distribution function (CDF); therefore, the task of locating a key reduces to \emph{learning the CDF of the sorted key set}. 
In a CDF plot, the $X$-axis represents the keys, and the $Y$-axis represents the rank of the keys. 
To approximate the entire key distribution, LIS introduces the \emph{Recursive Model Index} (RMI), a tree-structured, multi-stage architecture where models at a higher stage direct the queries to models at a lower stage to fine-tune the precision of the predicted key location. 

\begin{figure}[t]
    \centering
\includegraphics[width=0.42\textwidth]{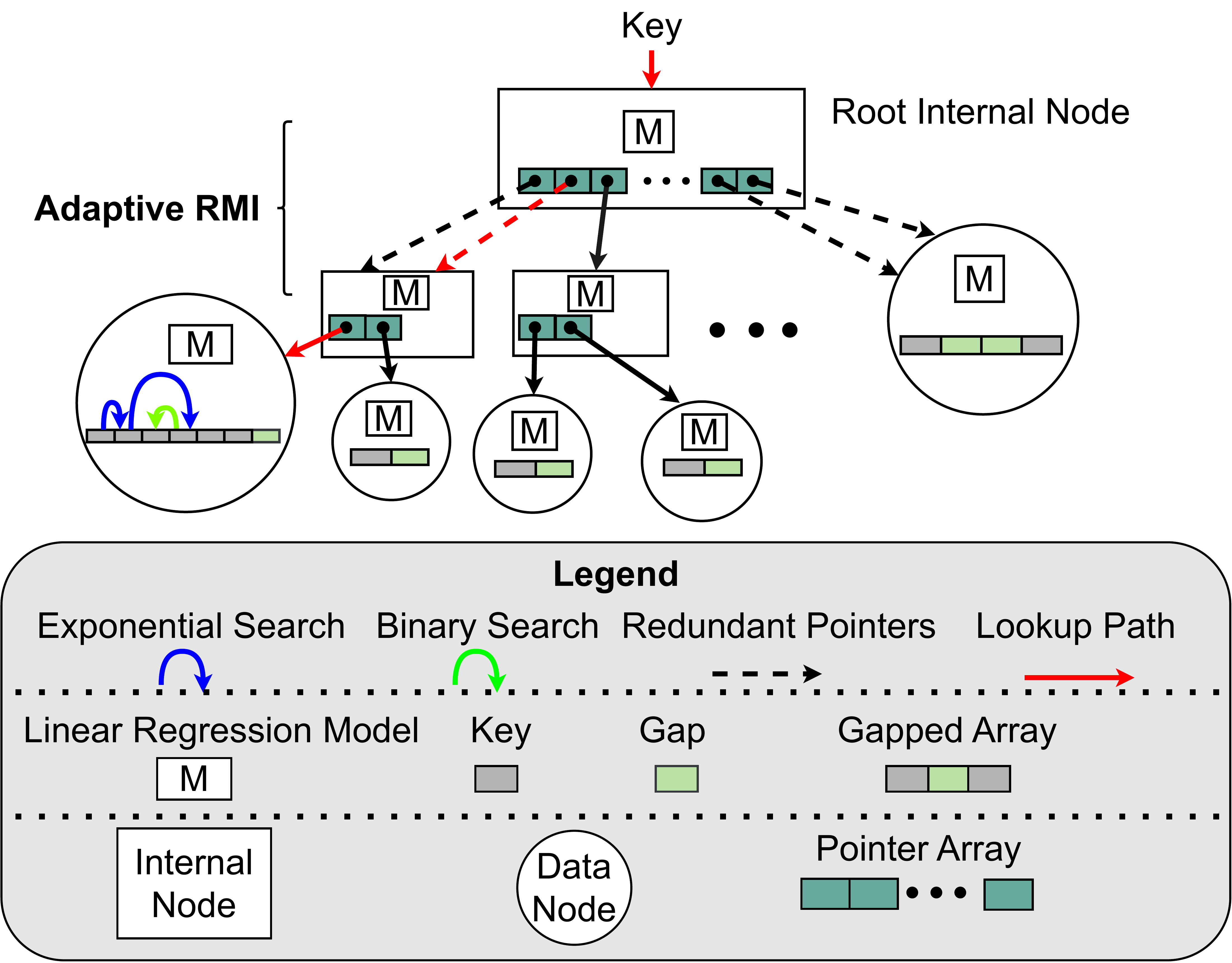}
    \vspace{-5pt}
    \caption{An illustration of ALEX. 
    }
    \label{fig:alex}
     \vspace{-2pt}
\end{figure}

\vspace{-2pt}
\subsection{ALEX: Dynamic Learned Index Structure}
\label{sec:prelim_alex}

A limitation of LIS~\cite{lis_sigmod18} is that it only supports lookup operations on read-only data. 
Thus, it is not applicable to scenarios where write operations are needed. 
A follow-up system called \emph{ALEX}~\cite{alex_sigmod20} was the first \emph{dynamic} learned index structure that extends the idea of LIS.  
At ALEX's core is a dynamic, tree-based index structure that consists of, much like the commonly-used B+tree structures~\cite{dbms_book}, a hierarchy of two types of nodes, namely ($i$) \emph{internal nodes} and ($ii$) \emph{data nodes}. 
Contrary to B+tree structures, each ALEX node is accompanied by a \emph{learning model} (i.e., linear regression) that, as we explain in the following, serves different functionality depending on whether it is a data node or an internal node. 

\noindent\textbf{The Role of Internal Nodes. }An internal node stores a linear model and a \emph{pointer array} that points to its children nodes. 
On a high level, the model takes as an input a key and outputs a location of its pointer array. 
Each entry of the pointer array is a reference to a model from the next level of the tree. 
We emphasize that all the entries of the pointer array point to child nodes; in case the array has more slots than the number of child nodes, then ALEX allows multiplicities for pointers, also called \emph{redundant pointers} (see Figure~\ref{fig:alex}). 
The following process generates redundant pointers:
Suppose that a node $X$ is split (a mechanism that is described in detail later); if $X$'s parent node has a single pointer in its pointer array associated with $X$, then the pointer array of $X$'s parent needs to be doubled to accommodate the two new nodes from $X$'s split. 
All the previous pointers are transferred to the new doubled-size array, and each pointer doubles its entries in the new array; that is if the sibling $Y$ of $X$ had 4 pointers pointing at it, in the new pointer array, there are 8 pointers that point to $Y$. 
The relation between the internal node and its children is the following: internal node $X$ is responsible for handling lookups/updates on a partition $[\alpha,\beta]$ of the key space, while each of the children nodes of $X$ is responsible for a non-overlapping (fine-grained) partition of $[\alpha,\beta]$; the role of $X$ is to propagate a lookup/update to the right partition. 
At the last level of the tree, partitions are associated with a data node that hosts keys that are distributed ``roughly'' linearly in the associated partition (see Figure~\ref{fig:alex}). 
In summary, working in conjunction with the data nodes, internal nodes serve two critical functions: 
(1)~Internal nodes direct queries from top to bottom to the right data nodes;
(2)~Internal nodes enable requests to be progressively directed to fine-grained (and variable size) partitions of the keyspace. 

\noindent\textbf{The Role of Data Nodes. }A data node stores a linear model and a \emph{gapped array} to store data elements. 
The empty spaces of the gapped array, i.e., the ``gaps'', are initially distributed uniformly during the node creation.
ALEX relies on these gaps to efficiently absorb new insertions and respond to lookups. 
The \emph{density} is defined as the ratio between the current number of keys in the gapped array and its total capacity. 
Each data node maintains the fraction of gaps between a fixed \emph{lower density limit} $d_l$ and a fixed \emph{upper density limit} $d_h$, which are (by default) set to $0.6$ and $0.8$, respectively.

\noindent\textbf{Lookups. }ALEX locates a key by performing a tree traversal from the top of the tree to the bottom. 
The query is directed from a parent internal node's linear model to one of its children/nodes until the query reaches a data node, also known as a leaf node in the context of a B+tree. 
Once the traversal reaches the correct data node, its linear model predicts the key position in the gapped array. 
During lookups, ALEX returns the key, if found,  in the predicted position; otherwise, there is a prediction error and ALEX performs an exponential search on the gapped array to find the key.

\noindent\textbf{Insertions. }For insertions, ALEX directly inserts the key into the predicted position; in case an existing key already occupies the predicted position, ALEX shifts existing keys toward the direction of the closest gap to create an empty slot for the new key. 
ALEX guarantees accurate placement of data by using \emph{model-based insertions} to ensure that:
(1)~for insertions, the input key is always inserted into the model-predicted position,
and (2)~for lookups, records are located close to the predicted position when possible. 

\noindent\textbf{Adapting to Workloads. }As the number of gaps decreases in a data node, the performance of both insertions and lookups degrades. 
To provide a good average-case performance, ALEX dynamically expands or splits data nodes to adapt to workload changes. 
ALEX does not wait until a data node becomes $100\%$ full; instead, ALEX \emph{expands} a data node's gapped array when the used space increase beyond the chosen upper-density limit. 
ALEX uses linear cost models to determine if, after the updates, the trained model of a data node deviates from the true distribution of keys. 
The cost model uses simple heuristics to compare (1) the expected (modeled) cost computed at the data node creation time to (2) the real-time empirical cost calculated by counting the number of exponential searches (for mispredicted lookups) and memory shifts (for insertions that cause key shifts) that have occurred since data node's creation. 

\noindent\textbf{Handling Corner Cases. }In addition, ALEX introduces heuristics to handle edge cases when a data node's cost increases beyond a threshold. 
Such edge cases are deemed as ``catastrophic'' events that significantly deteriorate ALEX's performance.
In these cases, ALEX \emph{splits} the data node affected by a catastrophic event to reduce the cost. 
ALEX implements two kinds of splits: \emph{splitting sideways} and \emph{splitting downwards}. 
A sideways split, which is the more common case,
directly partitions a data node's key space in half and assign the first half of redundant pointers (if any) to the left child data node and the second half of redundant pointers to the right child data node. 
If the parent internal node runs out of redundant pointers, ALEX doubles the size of the parent node's pointer array and adjusts its linear model parameters (the intercept and the slope) accordingly. 
Downwards splits only happen when splitting sideways is no longer possible. 
That is, the pointer array of the parent internal node is $16$MB (can be tuned at initialization).

\vspace{-2pt}
\subsection{Algorithmic Complexity Attacks}

In ACAs~\cite{aca_usenixsec03}, an attacker uses a small number of adversarial inputs to introduce a disproportional amount of work to the target system, pushing the system into using all available resources. 
ACAs are tailored to the algorithmic approach at hand and force a worst-case performance for the system. 
ACAs can be categorized into \emph{time ACA} where an attacker aims to exhaust the CPU resources~\cite{surgeprotector_sigcomm22, perffuzz_issta18, asawnf_sigcomm18, slowfuzz_ccs17}, and \emph{space ACA}, where an attacker aims to exhaust the memory or disk resources~\cite{spaceaca_journalacm81}.  
Take the hash-table attacks~\cite{aca_usenixsec03} as a time ACA example. 
A hash table can degrade to a linked list with carefully chosen inputs whose hash values map to the same bucket.   
Decompression bombs~\cite{aca_intro} exploit the capability of efficient compression algorithms to mount a space ACA by decompressing files and quickly consuming a large amount of storage space. 

ML for systems is gaining traction and is becoming an increasingly powerful tool for improving modern computer systems~\cite{lis_sigmod18, db_autotune_sigmod17, selfdrivingdb_cidr17, decima_sigcomm19, relaxed_belady_nsdi20, rl_mgmt_hotnets16}. However, incorporating ML into complex computer systems inevitably expands the attack surface, leading to increasing vulnerabilities. 
Poisoning attacks~\cite{clustering_adversarial_aisec13, feature_poisoning_icml15, yang2017generative, poisoning_byzantine_sec20} consider adversaries that deliberately augment the training data to manipulate the results of ML models. 
Kornaropoulos~et~al.~\cite{poisoning_lis_sigmod22} were the first to demonstrate poisoning attacks (which can be thought of as a time ACA) against the original, static LIS. 
However, when it comes to dynamic LIS like ALEX, the impact of injecting poisoning keys is not well understood.  
This is primarily due to ALEX's ability to dynamically adapt its key space partitioning and retrain its models, which helps mitigate the effects of the poisoning attack.

Similar to data structures that are vulnerable to ACAs, ML systems commonly employ ML models that are trained to optimize average-case objectives. However, these ML systems often lack 
robust guarantees regarding worst-case complexity. As we demonstrate in this work, an attacker can exploit ALEX's worst-case complexity---node splits or expansions---by inserting deliberately chosen inputs to mount space and time ACAs.

\vspace{-2pt}
\subsection{Threat Model}
\label{sec:threat_model} 

\noindent\textbf{Attacker’s Goal. }
ALEX~\cite{alex_sigmod20} offers the ability to adjust its shape to a dynamic workload adaptively. 
In this work, our focus is on adversaries who insert maliciously chosen keys into ALEX to trigger worst-case behaviors, ultimately using all available machine resources.  
The motives behind such adversarial behaviors vary depending on the potential gains and the context of the application, such as a competitor seeking to degrade performance or a denial-of-service attack. Specifically, in \cref{sec:data_node_aca} and \cref{sec:internal_node_aca}, our objective for space ACAs is to use all host's memory. 
In \cref{sec:time_aca}, the time ACAs aim to degrade the throughput of ALEX.

\noindent\textbf{Attacker’s Knowledge.}
In this work we study white-box, gray-box, and black-box attacks. 
In the white-box scenario, the attacker has full access to the inner-working of the ALEX under attack, including parameters of the linear regression models and internal structures (data nodes, internal nodes, and keys).
White-box attacks~\cite{aca_usenixsec03, fs_aca_sp09, surgeprotector_sigcomm22, tse_arxiv20, tse_conext19} are a natural first step towards assessing the robustness of a target system since they \emph{quantify the maximum damage}, i.e., act as an upper bound to any future black-box attack, which an adversary can make to a system. 
Previously proposed ACAs have primarily been developed in the white-box setting. 
For example, in~\cite{aca_usenixsec03}, the adversary has full knowledge of the targeted hash table (or binary tree) in all three scenarios examined, namely ACAs against Perl's hash tables, Squid web proxy, and Bro intrusion detection.  
Similarly, in~\cite{tse_arxiv20, tse_conext19}, the adversary has full knowledge of the control rules of Open vSwitch (OVS) so as to mount an ACA. 
 
\noindent\textbf{Attacker's Capabilities. }
We assume that the attacker can perform insertions into the target ALEX instance. 
There are several scenarios where an attacker can insert data to a target DB.
First, shared DBs are extensively used in many application scenarios (but not yet using the LIS paradigm). 
Companies such as Meta deploy database systems that are shared by multiple applications~\cite{rocksdb_fast20}, e.g., UDB, ZippyDB, and UP2X.
Another example is Google's DB infrastructure, including Bigtable~\cite{bigtable_osdi06}, together with the learned index system built atop Bigtable~\cite{bigtable_lis}, serves many concurrent users and applications. 
If the above systems used an LIS, malicious users could potentially mount ACAs to target shared DB services. 
Furthermore, the YCSB benchmark is synthesized based on the workload patterns of real-world, multi-tenant database systems~\cite{ycsb_socc10}. 
Second, crowd-sourced databases~\cite{crowddb_sigmod11, csdb1_vldb11, csdb2_uist11, csdb3_vldb12} allow users to contribute their data.  
In such a setting, an attacker can contribute to these datasets so as to poison/attack the learned index. 
Even if the attacker does not have access to the exact dataset, it is enough to have knowledge of the underlying distribution in order to devise an approach for generating maliciously crafted keys (e.g., the black-box setting of our space ACA on data nodes).  
Many real-world applications use crowd-sourced databases, e.g. OpenStreetMap~\cite{osm_aws} and Amazon’s Mechanical Turk (AMT)~\cite{amt_aws}.

\noindent\textbf{Terminology.} For space ACAs, 
we denote the number of adversarial insertions as $a$ and the number of legitimate insertions as $l$. 
We test our attacks on different \emph{adversarial budget} scenarios between $5\%$ to $30\%$, i.e., different parameterization of $100\cdot(a/(a+l))$. 
For time ACAs,
we use $a$ to denote the number of adversarial insertions, $l$ to denote the number of legitimate insertions, and $b$ to denote the lookup operations on the legitimate keys. 
In this case, the adversarial budget definition changes to $100\cdot(a/(a+l+b))$. 
We evaluate the effect of adversarial insertions under two workloads: $(1)$~write-heavy and $(2)$~read-heavy. 
The write-heavy workload consists of $50\%$ inserts and $50\%$ lookups, 
while the read-heavy workload has a ratio of $10\%$ inserts and $90\%$ lookups. 
\vspace{-2pt}
\section{Algorithmic Complexity Attacks Against ALEX Index Structure}
\label{sec:attacks}
In this section, we present three ACAs that aim to manipulate the resource utilization of the ALEX. Two of these attacks focus on memory consumption, while the third targets CPU usage.
We begin by introducing a space ACA that focuses on data nodes in~\cref{sec:data_node_aca}. 
Next, in~\cref{sec:internal_node_aca}, we present a space ACA that exploits the structure of \emph{internal nodes} of ALEX.   
Finally, in~\cref{sec:time_aca}, we present a time ACA that degrades the throughput of  ALEX. 
Our attack exposition is organized as follows: (1) presenting the ALEX design choice that the attack relies on, (2) the attack method, and (3) evaluation. 

\vspace{-2pt}
\subsection{Space ACA on Data Nodes}
\label{sec:data_node_aca}
In this subsection, we propose our first space ACA, which targets the memory utilization of the host system through exploiting the data node over-provision logic. 
Our experiments show that our attack uses up to $30\%$ more memory than ALEX~\cite{alex_sigmod20} compared to using legitimate keys. 
We tested the attack on all four datasets used in the original ALEX paper~\cite{alex_sigmod20} and reported the results. 
\vspace{-2pt}
\subsubsection{Exploitable Design Choice} 
\label{sec:data_node_vul}
ALEX relies on \emph{model-based insertion} (see ``Insertions'' paragraph in~\cref{sec:prelim_alex}) to decide a memory location for an inserted key. 
The driving force of model-based insertion is the gapped array, a data structure that balances the efficiency/fragmentation trade-off in this dynamic setting. 

\noindent\textbf{Trade-offs in Gapped Array. }
The gapped array prioritizes the speed of insertion over memory usage. 
Specifically, due to the model-based insertion, the gapped array guarantees efficient insertion with an $O(logN)$ time complexity (see~\cite{alex_sigmod20} for the analysis). 
However, the way of inserting (i.e., forcefully placing the key in the predicted location by shifting existing keys to make space) introduces long consecutive segments of allocated keys.
As a result, in case of a lookup of a shifted key, ALEX's original prediction is wrong, which triggers a ``corrective'' search mechanism that locates the shifted key. 
To compensate for slower lookups caused by long consecutive segments, ALEX \emph{preemptively migrates} to a larger gapped array before it reaches $100\%$ capacity. 
Specifically, ALEX introduces lower and upper density limits, $d_l=0.6$ and $d_h=0.8$, for each gapped array. 
A data node is considered \emph{full} if it reaches the upper limit $d_h$. 
Once a data node is full, ALEX performs either a node expansion or a node split, depending on the calculation of the data node's cost model (\cref{sec:internal_node_aca}). 
After the above action, ALEX increases the capacity of the new data node(s) to $\frac{\# \text{ records}}{d_l}$. 

\noindent\textbf{On Exploiting the Over-Provisioning Logic. }Notice that given the described over-provisioning approach, the empty space in a gapped can be at most $40\%$ (which occurs when being at $d_l$ capacity, i.e., right after a split/expansion) and at least $20\%$ (which occurs when being at $d_h$ capacity, i.e., before a split/expansion). 
We make the following observation: 
If the attacker constructs an insertion sequence that pushes a gapped array to $d_h=0.8$ density, then the very next insertion to this data node will trigger a split/expansion which will initialize the new gapped with $40\%$ empty space. 
We study the problem of how to craft such insertion sequences and  maximize the memory allocation by exploiting the over-provisioning logic. 
Interestingly, since each split/expansion increases the empty space \emph{proportionally} to the data node's relative size, we can see that some data nodes may give us more empty space than others. 
Thus, the attacker has to choose which data node to target first. 
\vspace{-2pt}
\subsubsection{Attack Method}
An attacker's goal is to maximize the memory usage of ALEX, given a limited budget of adversarially chosen keys.
A greedy strategy is to process the data nodes in descending order of size, and add as many keys as needed to each processed node so as to cause a split/expansion, continuing until the budget is exhausted. 
However, such a greedy approach is not optimal.

\noindent\textbf{Why the Greedy Approach Does Not Work? }
Consider the case where the largest data node requires $X$ keys to reach $d_h$ density and perform a split/expansion. 
Let $Y$ be the number of keys corresponding to the available budget, if $Y$ is smaller than $X$, then the greedy approach will ``waste'' the entire budget without causing a single split/expansion. 
This example signals that a different algorithmic approach is needed to optimaly allocate the available budget.

\noindent\textbf{Towards a Knapsack Formulation.}
In the following, we detail how the attacker's choice of which data node to target to use the available budget better can be seen as a knapsack problem (KP). 
Knapsack is a fundamental problem in combinatorial optimization with numerous applications to resource allocation scenarios. 
In KP, there are $n$ items, each of which with value $v_i$ and weight $w_i$, and the objective is to choose a subset $S$ of items so that their value is as high as possible while their overall weight is less than a given upper-bound $W$. 
In more technical terms, in this work, we are interested in the \emph{0-1 Knapsack Problem} where each item $i$ is associated with a variable $x_i\in\{0,1\}$ which takes value $1$ if the $i$-th object is chosen as part of $S$ and $0$ otherwise. 

Towards mapping the above terminology to our space ACA, a data node can be seen as an item. 
Let $i\in[1,n]$ be an indexing of the data nodes. 
The ``weight'' of a data node corresponds to the number of keys $k_i$ that are required to be inserted until the next expansion/split. 
The ``value'' of the data node corresponds to the free space (in bytes) $f_i$ of the data node's gapped array after the next expansion/split. 
Finally, the ``budget'' is the number of keys $B$ the attacker can insert. 
In this formulation, the goal is to pick a subset $S$ of data nodes, denoted by assigning $\{x_i=1\}_{i\in S}$ and $\{x_i=0\}_{i\notin S}$, so as to maximize the objective function $\sum_{i\in [1,n]}x_i\cdot f_i$ while satisfying the constraint $\sum_{i\in [1,n]}x_i\cdot k_i< B$ and $x_i\in\{0,1\}$.  
Notice that the solution from the above formulation tells the attacker \emph{how many} keys need to be allocated to each data node but not \emph{which} keys to insert. 
Recall that each data node is responsible for a consecutive range of the key domain. 
Therefore, when the attacker wants to generate $k_i$ keys to be inserted into a data node, (s)he can simply choose uniformly at random among the corresponding range. 

\noindent\textbf{Extending to Multiple-Choice Knapsack. }
The above formulation of 0-1 KP is a good start but it is missing a large number of potential strategies for the attacker. 
Specifically, once the $i$-th data node is chosen (by assigning $x_i=1$), the aforementioned formulation gives no option to consider expanding the same data node again. 
Thus, in case of an instance where optimal space ACA is given by extending/splitting the same data node multiple times, the above formulation will give a suboptimal solution.

In the following, we present how to extend our KP formulation to account for such scenarios. 
Multiple-Choice Knapsack (MCK) is a variant of KP where the items are subdivided into $m$ classes. 
MCK has all the previously described constraints of KP plus an additional constraint that ensures that exactly one item is chosen from each class of the $m$ classes.
The main insight for extending our formulation is to consider each data node as a class of its own and generate $m$ distinct scenarios. 
In each scenario of a data node, the inserted keys cause a different number of expansions/splits. 
Thus, we \emph{hard-code} additional scenarios, each describing the same data node being targeted multiple times.  
The constraint that ``only one out of $m$ choices is allowed'' guarantees that we won't choose the same data node to expand, for example, both once and twice. 
The last step to reach our final formulation is to introduce a scenario, the $(m+1)$-th one, that allows a data node not to be chosen at all. 
To capture the MCK notation, we introduce the sub-index $j$ that takes values from $1$ to $m+1$ and denotes which scenario is chosen for the corresponding data node. 
Thus, $x_{ij}$ indicates whether the $j$-th scenario of the $i$-th data node is chosen, $f_{ij}$ indicates the free space that the attacker generates by targeting the $i$-th data node to be expanded/split multiple times according to the $j$-th scenario, and $k_{ij}$ indicates the number of keys that the attacker needs to insert to the $i$-th data node to be expanded/split multiple times according to the $j$-th scenario. 
The complete formulation is the following:
\vspace{-2pt}
\begin{equation}
\resizebox{.26\textwidth}{!}{%
$
\begin{aligned}
\texttt{max}_{x_{ij}} \sum\nolimits_{i=1}^{n}\sum\nolimits_{j=1}^{m+1}& x_{ij}\cdot f_{ij}\\
\text{s.t. } \sum\nolimits_{i=1}^{n}\sum\nolimits_{j=1}^{m+1} x_{ij}\cdot k_{ij} &\leq B,
\\
\sum\nolimits_{j=1}^{m+1} x_{ij}&=1, \forall i\in[1,n] \\
x_{ij} &\in\{0,1\}
\end{aligned}$
}
\label{eq:mck}
 \vspace{-2pt}
\end{equation}

\begin{figure*}[t]
\begin{center}
\subfigure[{\small\texttt{Longitude}}: 50M keys.] {
\includegraphics[width=.31\textwidth]{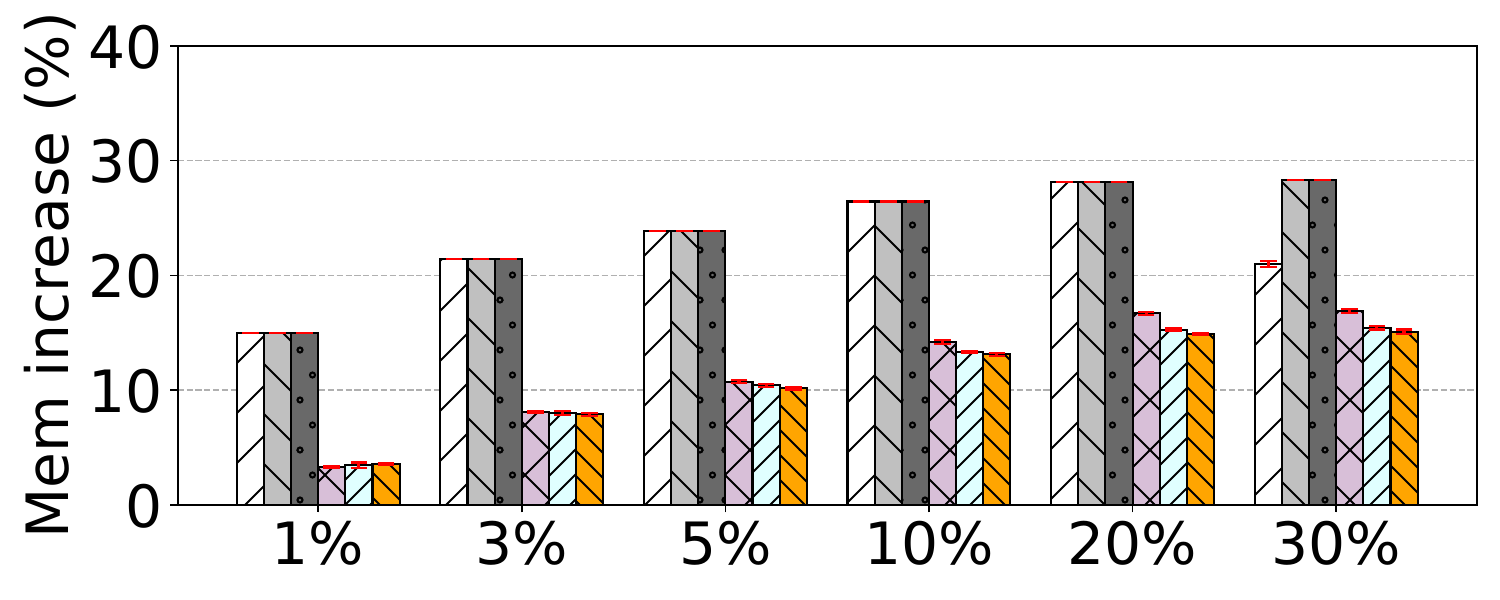}
\label{fig:dn_eval_longitudes_50M}
}
\hspace{-7pt}
\subfigure[{\small\texttt{Longitude}}: 100M keys.] {
\includegraphics[width=.31\textwidth]{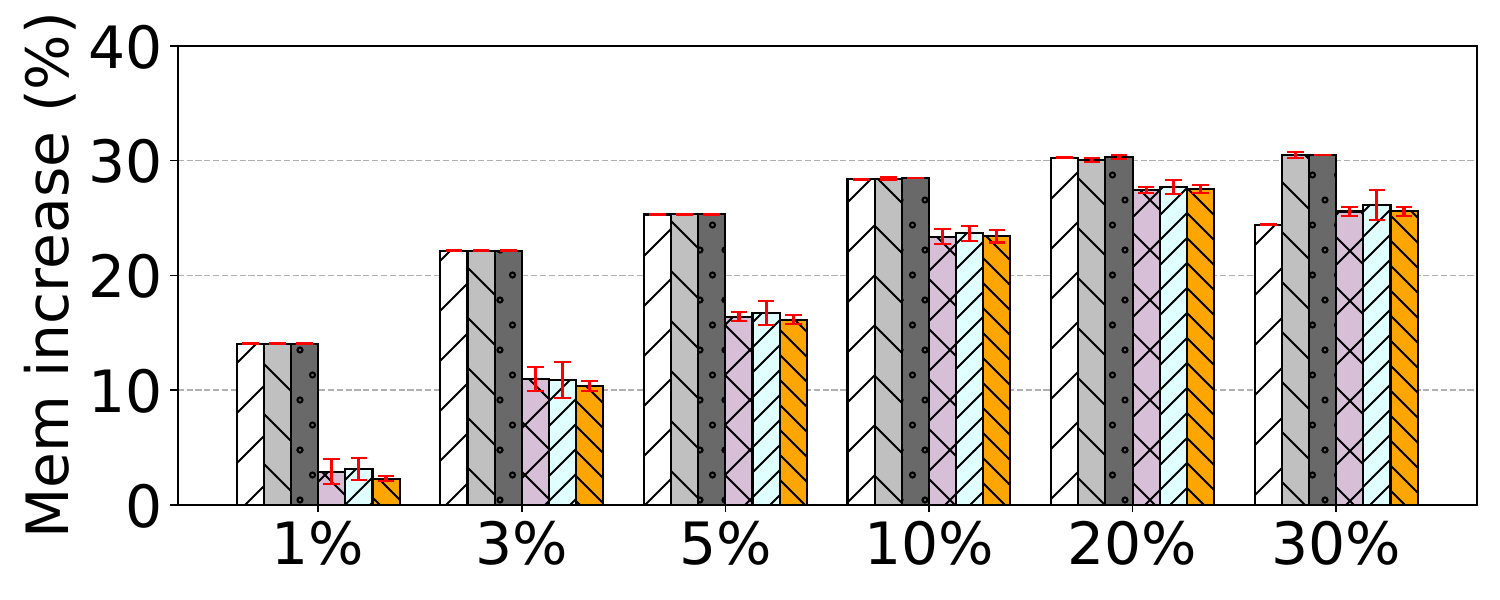}
\label{fig:dn_eval_longitudes_100M}
}
\hspace{-7pt}
\subfigure[{\small\texttt{Longitude}}: 150M keys.] {
\includegraphics[width=.31\textwidth]{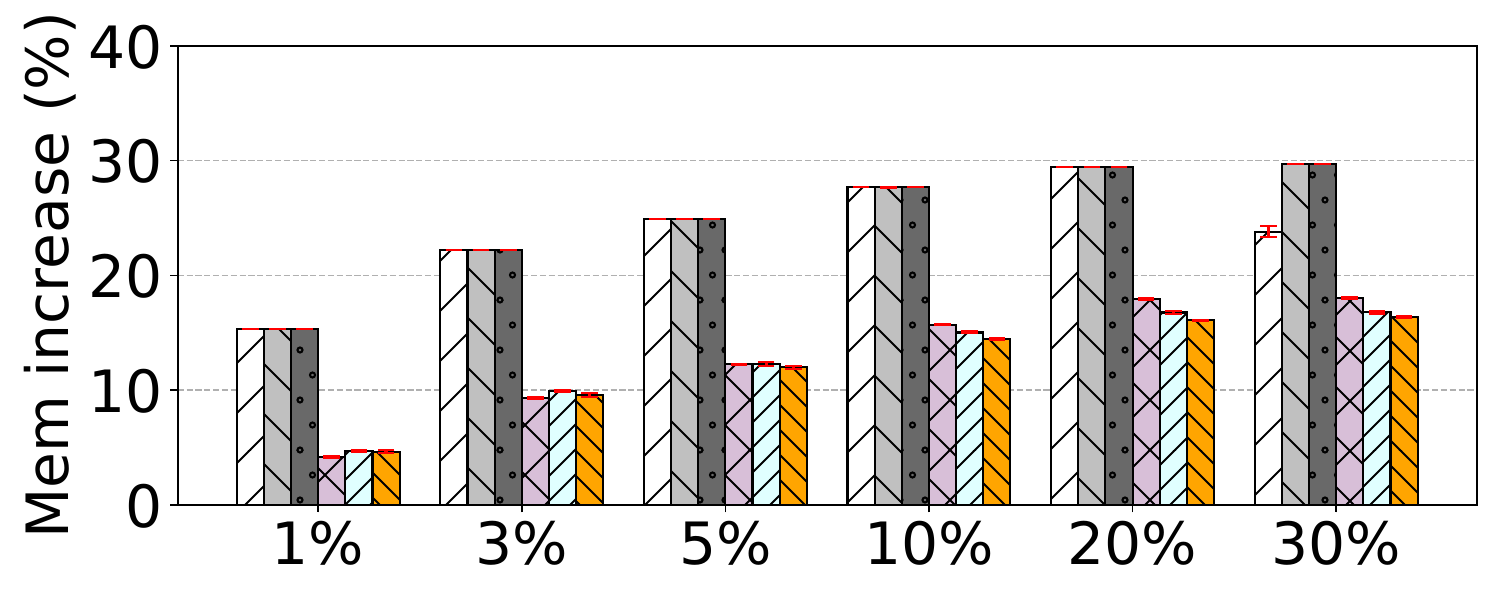}
\label{fig:dn_eval_longitudes_150M}
}
\vspace{-12pt}
\end{center}
\vspace{-10pt}
\end{figure*}

\begin{figure*}[t]
\begin{center}
\subfigure[{\small\texttt{Longlat}}: 50M keys.] {
\includegraphics[width=.31\textwidth]{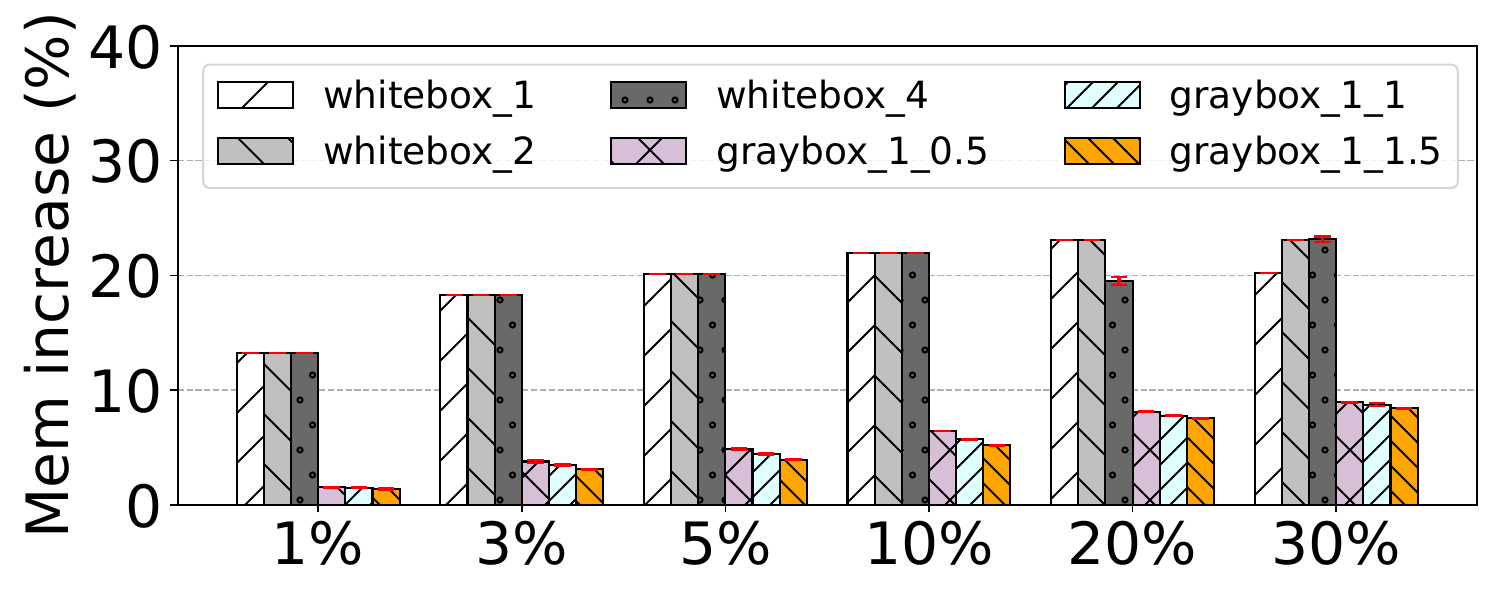}
\label{fig:dn_eval_longlat_50M}
}
\hspace{-7pt}
\subfigure[{\small\texttt{Longlat}}: 100M keys.] {
\includegraphics[width=.31\textwidth]{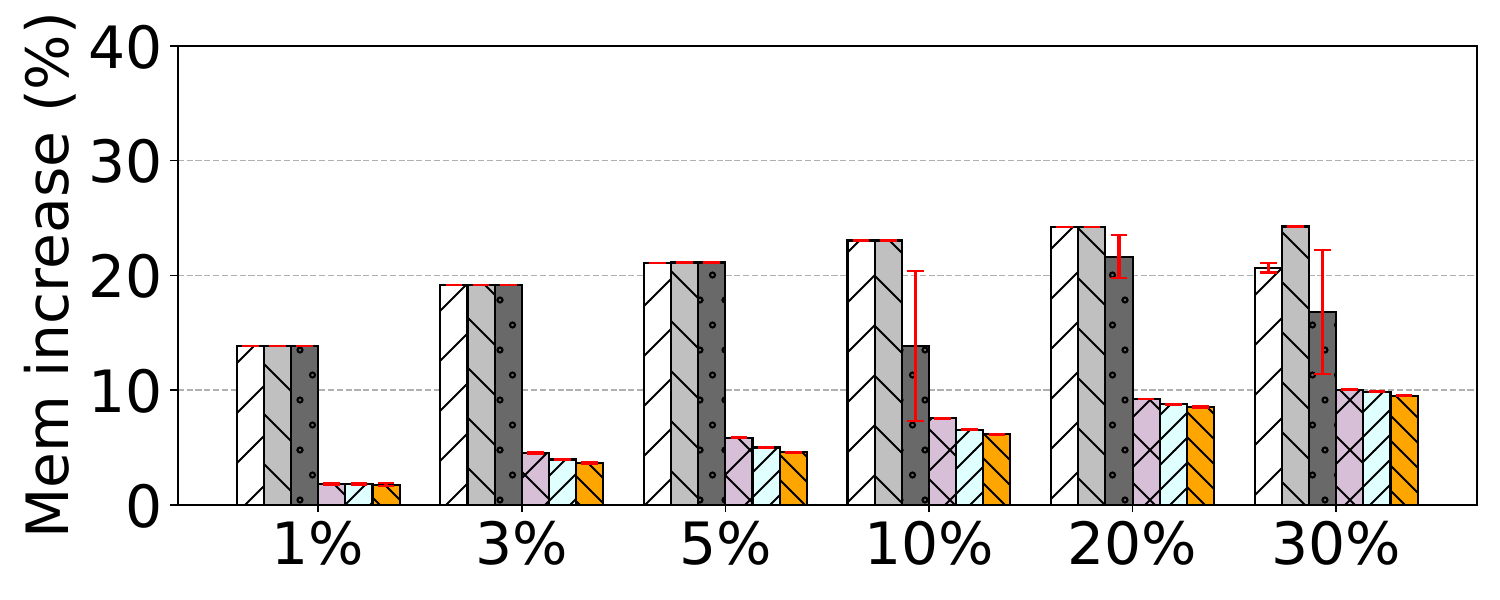}
\label{fig:dn_eval_longlat_100M}
}
\hspace{-7pt}
\subfigure[{\small\texttt{Longlat}}: 150M keys.] {
\includegraphics[width=.31\textwidth]{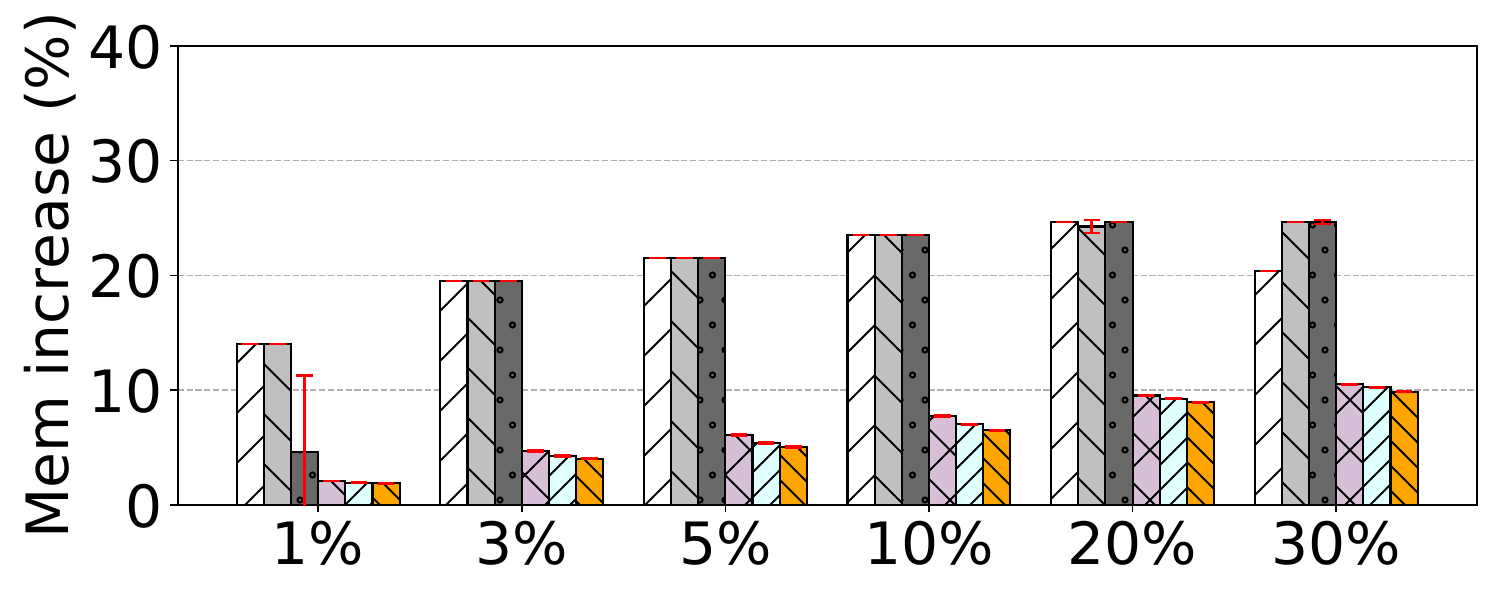}
}
\vspace{-12pt}
\end{center}
\vspace{-10pt}
\end{figure*}

\begin{figure*}[t]
\begin{center}
\subfigure[{\small\texttt{YCSB}}: 50M keys.] {
\includegraphics[width=.31\textwidth]{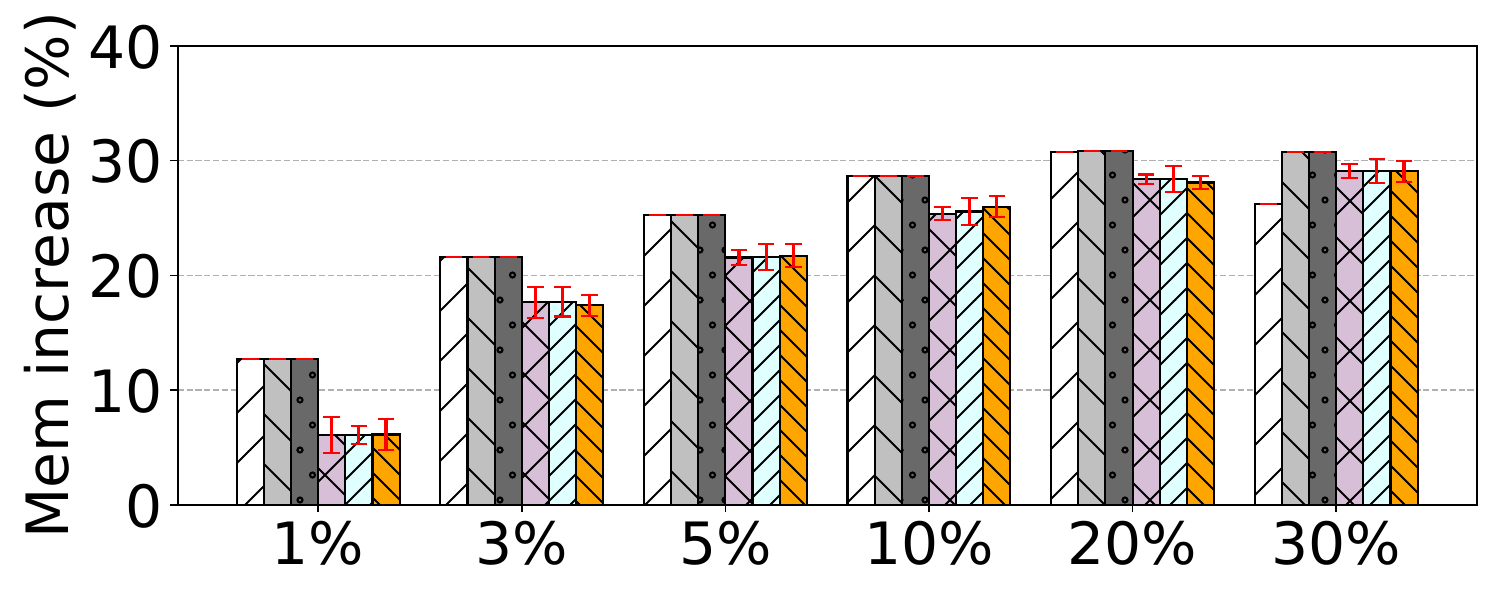}
\label{fig:kde_longitude}
}
\hspace{-7pt}
\subfigure[{\small\texttt{YCSB}}: 100M keys.] {
\includegraphics[width=.31\textwidth]{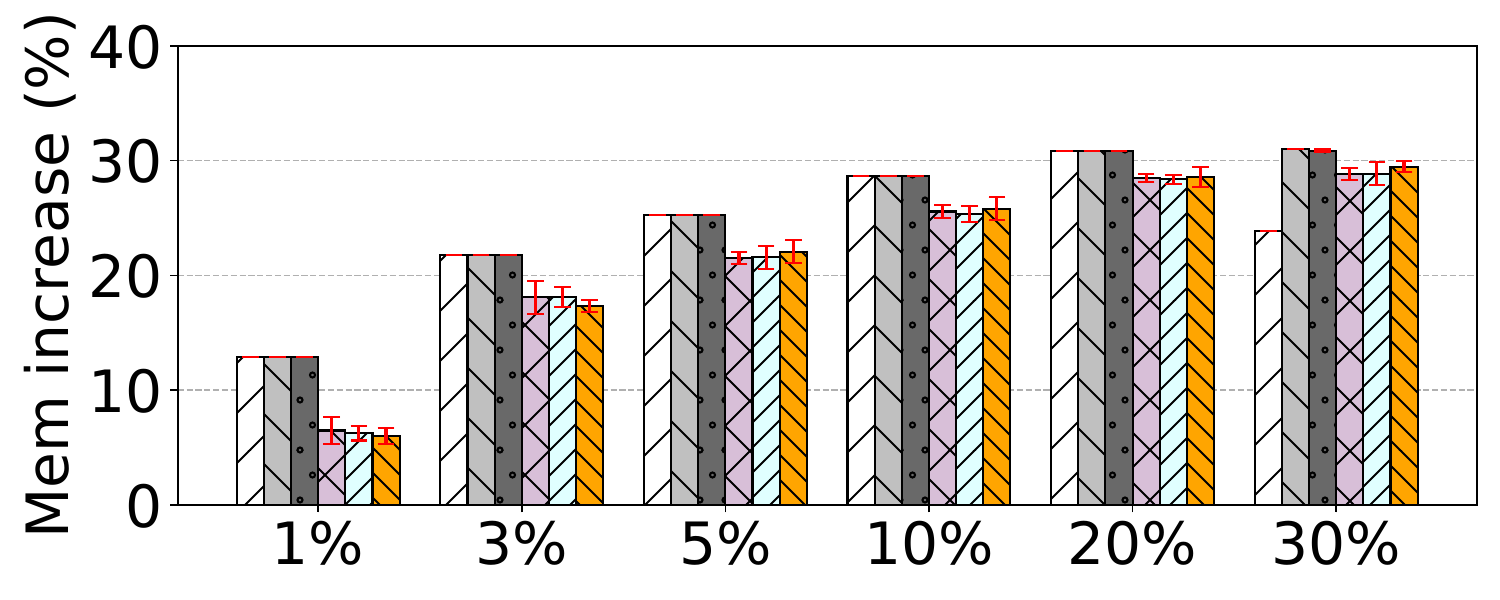}
\label{fig:kde_longlat}
}
\hspace{-7pt}
\subfigure[{\small\texttt{YCSB}}: 150M keys.] {
\includegraphics[width=.31\textwidth]{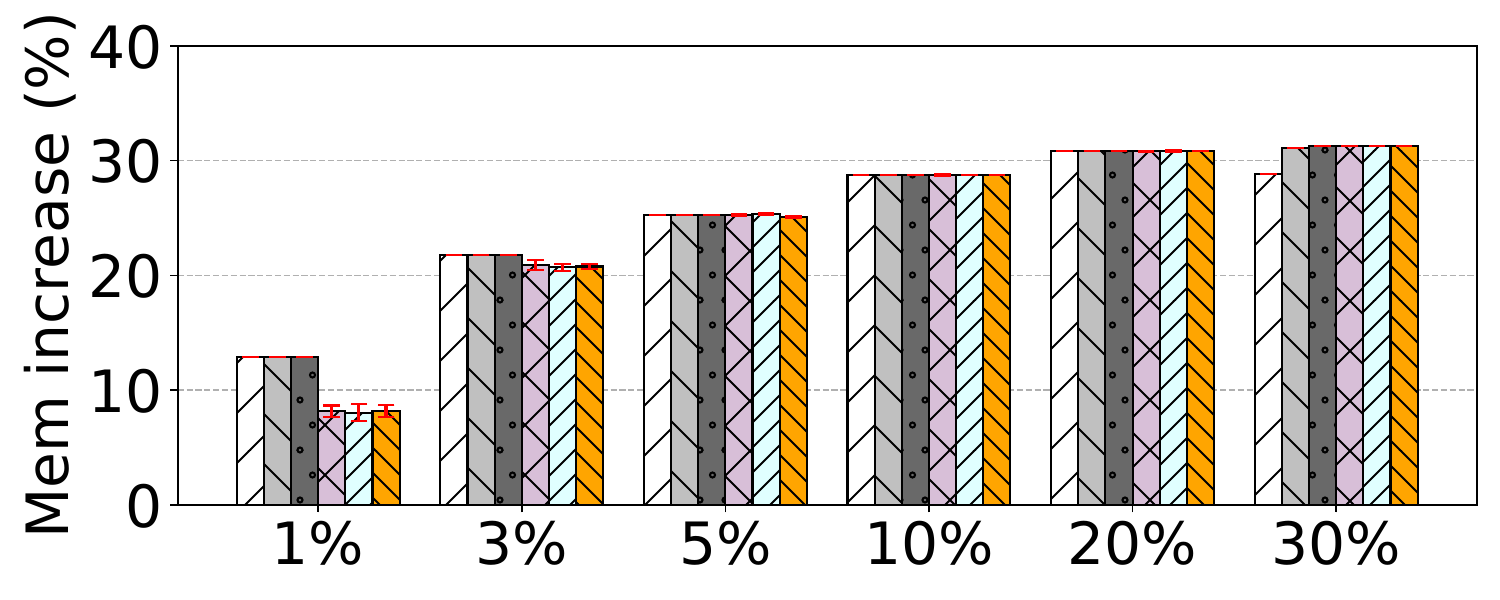}
\label{fig:dn_evals_ycsb_150m}
}

\vspace{-12pt}
\end{center}
\vspace{-10pt}
\end{figure*}

\begin{figure*}[t]
\begin{center}
\subfigure[{\small\texttt{Lognormal}}: 50M keys.] {
\includegraphics[width=.305\textwidth]{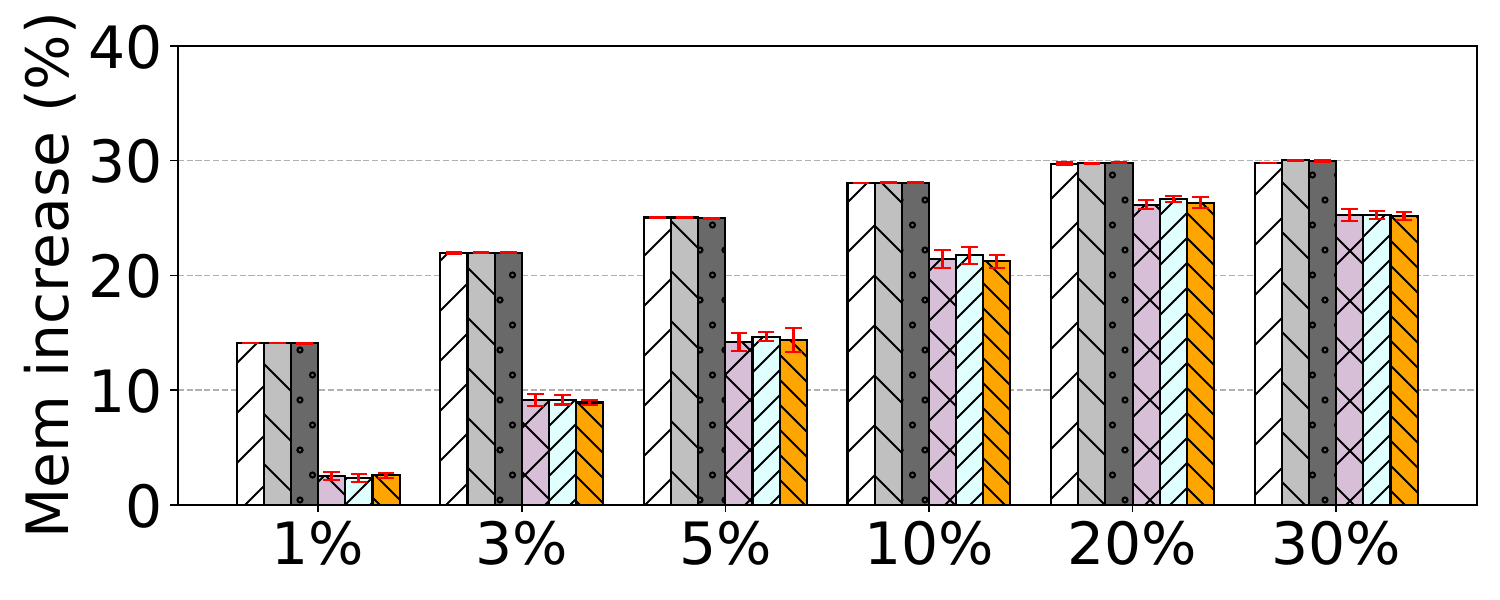}
}
\hspace{-7pt}
\subfigure[{\small\texttt{Lognormal}}: 100M keys.] {
\includegraphics[width=.305\textwidth]{plots/space_aca_dn/dn_lognorm_100m.pdf}
}
\hspace{-7pt}
\subfigure[{\small\texttt{Lognormal}}: 150M keys] {
\includegraphics[width=.305\textwidth]{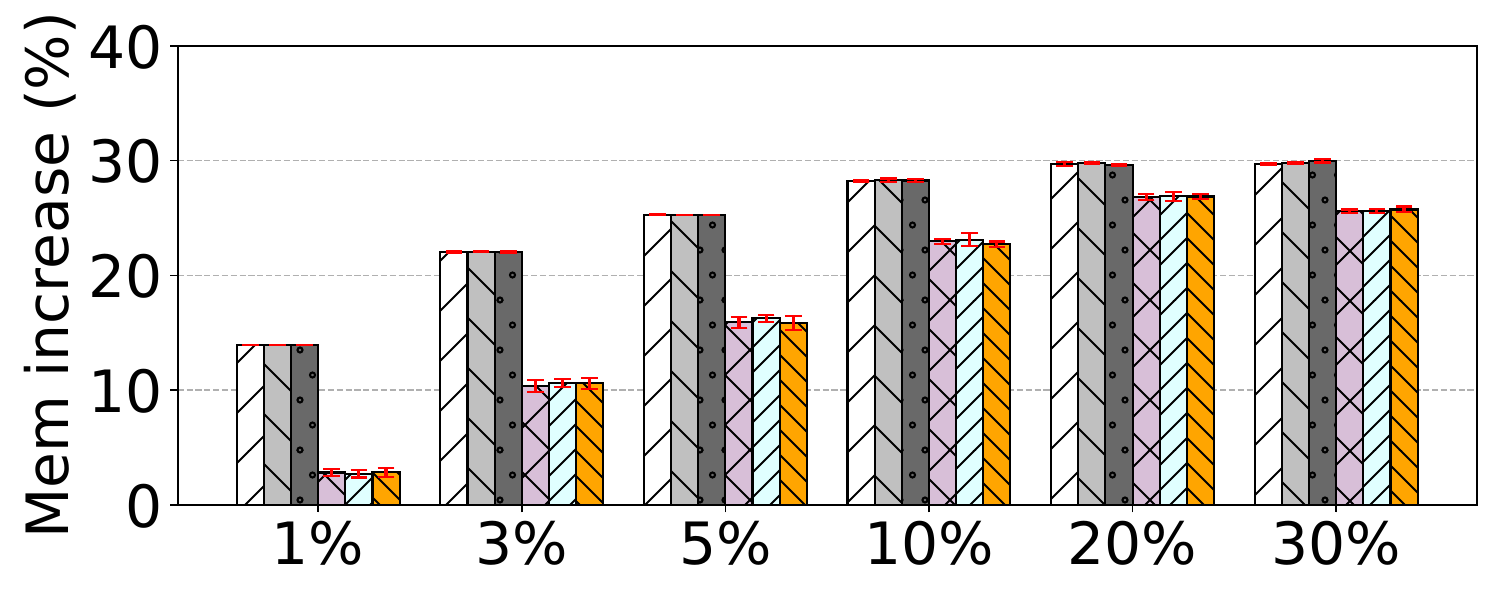}
}
\vspace{-12pt}
\caption{
Memory increases of space ACA on data nodes.~\textit{\textmd{This figure shows the normalized (with respect to the baseline) memory increase for white-box and gray-box attacks using four datasets and varying ALEX sizes of 50~million, 100~million, and 150~million. The X-axis is the percentage of attacker's budget with respect to the total keys ranging from 1\% to 30\%; the Y-axis is the normalized memory increase from our attacks. 
{\small\texttt{whitebox\_E}} denotes a white-box attack that allows at most $E$ expansion(s) or split(s) for any data node for the MCK optimization; 
for {\small\texttt{graybox\_E\_b}}, $E$ denotes a gray-box attack that allows at most $E$ expansion(s) or split(s) for any data node during the MCK optimization, and $b$ is the ``bandwidth'' parameter for KDE.
Each data point is the mean of five runs. The error bars are depicted in red.}}}
\label{fig:dn_evals}
\end{center}
\end{figure*}
\noindent\textbf{White-Box \& Gray-Box Extensions. }
The following details how the MCK formulation applies to white-box and gray-box attacks. 
In the white-box scenario, the attacker has access to all the data nodes of the target ALEX. 
Subsequently, the attacker uses an MCK solver to determine an offline plan of which data nodes to target and then proceeds to insert adversarial keys based on the generated plan. 
In the gray-box scenario, the attacker has only \emph{distributional knowledge} about the target ALEX key distribution. 
In this case, the attacker generates synthetic data using distributional knowledge and locally constructs a substitute ALEX index. 
Utilizing this substitute ALEX, the attacker generates a plan using the MCK solver and inserts the resulting sequence of keys to the target ALEX. 

\noindent\textbf{Discussion.}
For simplicity, in this work, we assume a gray-box setting from which the attacker has access to the \emph{key distribution} (modeled using a Kernel Density Estimation technique); but one can instead query the targeted ALEX instance in the black-box setting and \emph{learn an approximation of the key distribution}. 
After this step, the black-box attack proceeds exactly like the gray-box. 
That is, it samples the approximated key distribution to build a local substitute model; it then crafts an attack on the local model and applies the adversarial payload to the target model.

\vspace{-2pt}
\subsubsection{Evaluation}
\label{sec:data_node_eval}

We implement the MCK solver using Google's OR-Tools~\cite{tool_ortools}. 
For the substitute ALEX, we use \texttt{scikit-learn}'s Kernel Density Estimation library~\cite{tool_sklearn}. 
The \texttt{bandwidth} parameter for KDE takes values $\{0.5, 1, 1.5\}$ and  the \texttt{kernel} parameter is {\small\texttt{tophat}}. (Refer to \cite{tool_sklearn} for detailed information on the above parameters.) 
We tested our data-node space ACA on an EC2 {\small\texttt{m5.4xlarge}} VM instance with 16 vCPUs and 64GB memory. 
We tested the white-box and gray-box attacks five times and measure the mean  increase in memory consumption compared to the memory usage under a legitimate workload with the same number of keys (insertions).

\noindent\textbf{Datasets. }
We tested four datasets with 8-byte keys: {\small\texttt{Longitudes}} and {\small\texttt{Longlat}} use 8-byte {\small\texttt{double}}-precision floating pointer numbers as keys, while {\small\texttt{YCSB}} and {\small\texttt{Lognormal}} use 8-byte {\small\texttt{int}} as keys. 
The {\small\texttt{Longitude}} workload contains keys that represent the longitudes of locations drawn from the OpenStreeMaps public dataset~\cite{osm_aws}. The {\small\texttt{Longlat}} workload contains mixed keys that combine the longitude and latitude information from the OpenStreetMap dataset by applying the transformation $k=180 * \lfloor{longitude}\rfloor + latitude$ to each pair of the longitude and latitude value.
The {\small\texttt{YCSB}} workload~\cite{ycsb_socc10} generates keys that represent user IDs uniformed distributed across the whole 64-bit {\small\texttt{int}} range. The {\small\texttt{Lognormal}} workload generates keys that follow the log-normal distribution with $\mu=0$ and $\sigma=0$. 

\noindent\textbf{Tested Methods. }
We evaluated our attack by comparing its memory consumption to the memory footprint of a normal sequence of insertions. 
We used different workload sizes (50M keys, 100M keys, and 150M keys). 
We tested the following scenarios:
\begin{itemize}[noitemsep,leftmargin=*]
  \item\textbf{Baseline}: We construct an ALEX instance by inserting legitimate keys drawn from the original datasets until reaching a target workload size and measured the memory usage of ALEX. The memory usage of the baseline is not shown in Figure~\ref{fig:dn_evals} as all bars are normalized with respect to the baseline. 
  \item\textbf{White-box Attack}:
The white-box attack follows three steps:
(1)~initialize an ALEX with $l$ keys from legitimate keys of the original dataset;
(2)~run the MCK solver that takes the following information as input: 
 one array that details how much budget would be needed if one were to trigger $E$ splits/expansions (where the parameter $E$ takes values $0$, $2^0$, $2^1$, and $2^2$; thus, the $j$ from Eq.~(\ref{eq:mck}) considers four scenarios) for each of the $n$ data nodes given the current state of the ALEX instance; the other array details the free memory increase from triggering $E$ splits/expansions for each of the $n$  data nodes given the current state of the ALEX instance; 
and an fixed budget $B$; 
given the input, the MCK solver generates the attack plan;
(3)~perform the sequence of insertions based on the plan generated from the previous step.

  \item\textbf{Gray-box Attack}: The gray-box attack follows five steps:
(1)~Use a Kernel Density Estimation (KDE) model to approximate the distribution of the original dataset. 
The KDE model takes as input the original dataset, a ``bandwidth'' parameter, and a ``kernel'' parameter;
(2)~initialize an ALEX with $l$ keys from legitimate keys of the original dataset;
(3)~construct a substitute ALEX instance by sampling $l$ keys from the learned KDE model; 
(4)~run the MCK solver using the data node information from the instantiated substitute ALEX and budget $B$;
(5)~issue the sequence of insertions based on the plan generated from the previous step.
\end{itemize}

\noindent\textbf{Results. }
Figure~\ref{fig:dn_evals}
presents normalized memory increases for gray-box and white-box attacks with respect to the baseline. 
We observe that, across all four datasets for the three workload sizes, both white-box and gray-box attacks result in an increase in memory consumption.
Notably, under the white-box setting, the four datasets have a memory increase between $12.76\%$ (for {\small\texttt{YCSB}}) and $15.32\%$ (for {\small\texttt{YCSB}}) with only $1\%$ of budget size.
The memory gap between white-box attacks and gray-box attacks is particularly striking for {\small\texttt{Longitude}} and {\small\texttt{Longlat}}, while this gap is relatively smaller 
for {\small\texttt{YCSB}} and {\small\texttt{Lognormal}}; there is even no difference for {\small\texttt{YCSB}} 150M. 
(shown in Figure~\ref{fig:dn_evals_ycsb_150m}). 
This is because {\small\texttt{Longitude}} and {\small\texttt{Longlat}} are not accurately approximated by the given number of samples to the KDE, which affects the performance of the attack. 
On the other hand, the other two datasets are much more linear. E.g., {\small\texttt{YCSB}} dataset is highly linear, which results in ALEX using fewer data nodes to store keys. 

There are no obvious differences for all three expansion parameters (1, 2, and 4) for budget sizes
below ${30\%}$ under white-box attacks. 
This is because, when the budget is relatively small, one expansion is sufficient to generate an optimal solution. 
Whereas at least two expansions are needed for ${30\%}$ in order to maximize memory increases. 
{\small\texttt{white\_box\_2}} and {\small\texttt{white\_box\_4}} perform better than {\small\texttt{white\_box\_1}} for most of the cases.
However, there are rare cases  
where {\small\texttt{white\_box\_4}} performs worse than {\small\texttt{white\_box\_1}} and {\small\texttt{white\_box\_2}} as shown in Figure~\ref{fig:dn_eval_longlat_50M} and Figure~\ref{fig:dn_eval_longlat_100M}. 
This is likely due to the computational time constraint we configured for the MCK solver implemented using the OR-Tools: 
we set a maximum search time of 100 seconds to strike a balance between solution optimality and time efficiency. 
If either the optimal solution was found within the time limit, or the configured time period elapsed, whichever condition was met first, the MCK solver returned the computed attack plan. 
We believe that this uncommon event appeared due to the non-linearity nature of the {\small\texttt{Longlat}} dataset. 
This non-linearity resulted in the creation of a large number of data nodes, 
which required a longer computational time period to search for the optimal solution. 
The memory increases under gray-box attacks 
largely depend on how close the state of the substitute ALEX instance to the target ALEX instance is and the linearity of the original dataset's key distribution. 

In summary, white-box attacks achieved ${12.76\%}$ to ${33\%}$ memory consumption increases, depending on the configuration of the expansion parameter and the attacker's budget.
On the other hand, gray-box attacks are not as close to the damage made by white-box attacks because the substitute ALEX instance of a gray-box setting may deviate from the true ALEX instance.

\noindent\textbf{Potential Mitigation.} Instead of using extra ``gaps'' at data-node level to absorb inserts, one can potentially use a delta buffer to serve inserts. The granularity of the delta buffer could be index-level (e.g., PGM~\cite{pgm_vldb20}) or node-level (e.g., FITing-Tree~\cite{ftree_icmd19}) depending on specific goals. 
These buffers typically are allocated as fixed size and merged within them~\cite{pgm_vldb20} or with existing data~\cite{ftree_icmd19}. Once a merge process is triggered, the buffer will be emptied and reuse. 
Adding delta buffers is an intrusive design choice that might break certain existing properties of ALEX: while eliminating the gapped array design can potentially defend against our space ACA, one potential side effect is the extra memory copy overhead when merging buffered inserts into a node array, which will further affect the performance of ALEX.  
\vspace{-2pt}
\subsection{Space ACA on Internal Nodes}
\label{sec:internal_node_aca}
In this subsection, we present our internal node ACA that exhausts the  memory consumption of the host. 
This attack exploits multiple design choices of ALEX, which combined lead to a cascading storage increase. 
We show how the current design of ALEX is problematic when it comes to duplicate keys. 
We note that supporting duplicate keys is a necessary functionality in index structures. 

\vspace{-2pt}
\subsubsection{Exploitable Design Choice} 
\label{sec:internal_node_vul}
\begin{figure*}[h]
    \centering
    \includegraphics[width=0.85\textwidth]{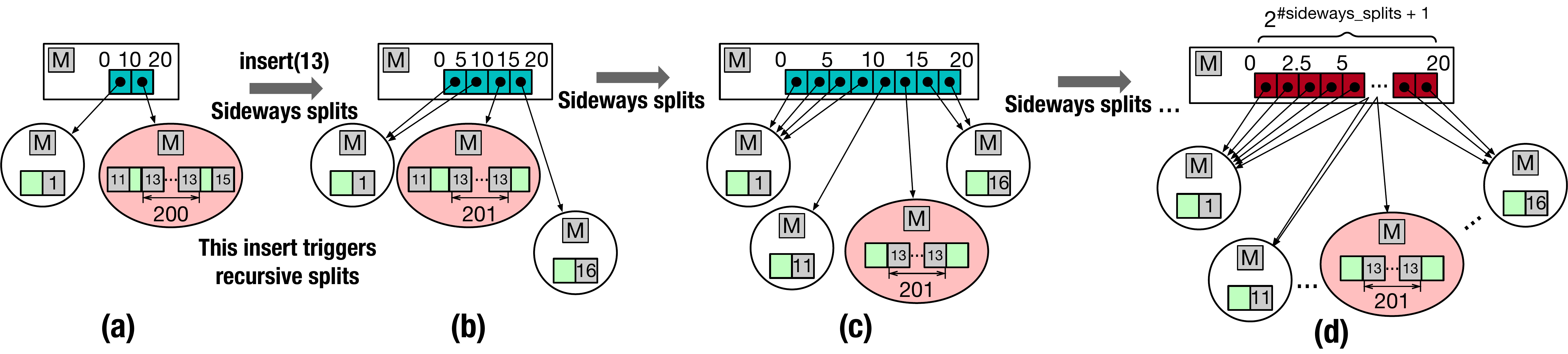}
    \vspace{-12pt}
    \caption{A Space ACA insertion of duplicate key 13 that causes recursive splits and ultimately results in an OOM error.  
    \textit{\textmd{(a)~The right data node (in light red) has accumulated 200 duplicate keys 13, exceeding the cost threshold; 
    (b)~Inserting one more key 13 triggers a sideways split;
    (c-d)~Attempting to lower the cost, ALEX recursively (cascadingly) sideways splits; the size of the internal node's pointer array keeps growing exponentially until reaching the maximum internal node size or ultimately causing an OOM; for the former case, ALEX performs a downwards split to create a new internal node at a lower level and then continues sideways splits; this cascading process continues until OOM.
    }}}
    \label{fig:ce}
\end{figure*}

Recall that ALEX performs insertions by traversing the tree downwards until it identifies the data node associated with the range to which the input key belongs. 
Our next attack uses several mechanisms of ALEX's insertion to cause \emph{infinite splits} using only a \emph{single} insertion.  
To achieve this devastating overhead, the attacker has to insert multiple copies of the same key, i.e., \emph{duplicate keys}. 
In the following, we unravel the series of events that cause an unbounded number of splits. 

\noindent\textbf{Impact of Inserting Duplicates on a Gapped Array. }
Suppose that a  data node already stores a key $\kappa$. 
Let's first analyze the case where $\kappa$ is inserted again. 
ALEX traverses the tree and lands on the same data node where the original $\kappa$ resides.
The linear model of the data node points to the entry in the gapped array where the ``older'' key $\kappa$ is stored. 
Since this position is occupied, ALEX shifts the old $\kappa$ to create a gap for the new $\kappa$. 
When the placement is done, the two $\kappa$ entries appear consecutively in the gapped array. 
The above description holds when the attacker inserts yet another copy of $\kappa$, in which case, ALEX shifts both of the older $\kappa$ entries to end up with three consecutive copies.  
The takeaway message is that  after a duplicate key insertion, the following conditions hold:
(1)~ALEX performs at least as many shifts in the gapped array as the number of identical keys already reside in it, and
(2)~duplicate keys are allocated consecutively. 

\noindent\textbf{Impact of Duplicates on the Cost Model. }
As a next step, we discuss how conditions (1) and (2) from above affect the cost model of the corresponding data node. 
The first thing worth pointing out is that periodically, the cost model assesses its accuracy (by calculating the \emph{cost}), and if the model deviates from the current set of keys, the cost is deemed too high, and ALEX triggers the so-called \emph{catastrophic event}. 
When this occurs, the data node is forced to perform a sideways split (as opposed to an expansion or a downward split). 
A sideways split of the data node ``redistributes'' the keys of the problematic data node to two new data nodes in the hope that each new data node will have a much lower cost.   
Unfortunately, conditions (1) and (2) from the previous paragraph both significantly affect the cost of their data node. 
The reason is that the number of shifts that have occurred so far is a crucial metric of the cost model; therefore, condition (1) exacerbates the cost on that front. 
In the same vein, long consecutive segments of keys in the gapped array, i.e., condition (2), increase the number of steps in the exponential search, which is the other crucial metric of the cost model.  
Thus, the hope is that the sideways split will redistribute the keys and reduce the cost.

\noindent\textbf{What Happens During a Split? }When ALEX needs to perform a sideways split, it will split the data node by equally partitioning its key space into half. 
We emphasize here that the word split does not refer to partitioning the set of keys of the data node in two equal-sized subsets but rather partitioning the \emph{key space} into two key ranges of equal length. 
This subtle but crucial detail is one of the main reasons the attack works so well. 
To be more precise, when a data node $Z$ is split, ALEX generates two new data nodes, e.g., $Z_L$ and $Z_R$, and each of them is responsible for half of $Z$'s key space. 
As a next step, ALEX relocates the keys of $Z$ that fall under the left key space to $Z_L$ and the keys that fall under the right key space to $Z_R$. 
We note here that this partition of keys can be unbalanced (in fact, our attack capitalizes on that possibility). 
As a final step, ALEX checks the parent of $Z$, denoted as $P_Z$,  to identify how many entries of $P_Z$'s pointer array are associated with $Z$.
If there is more than one pointer to $Z$
(an instance of the so-called \emph{redundant} pointers from Section \ref{sec:prelim_alex}) then half of them are redirected to point to $Z_L$ and the other half to $Z_R$. 
The interesting (for our attack) case occurs when $P_Z$ has only a single pointer to $Z$.
In that case, $P_Z$ is \emph{forced to double} the size of its pointer array. 
Thus, each entry of $P_Z$'s pointer array now occupies two entries in the new pointer array, which means that $P_Z$ can now assign one pointer to $Z_L$ and one to $Z_R$.

\noindent\textbf{Why Splits Do Not Reduce Cost from Duplicates? }
The Achilles' heel of this design is their strategy of halving the key space (and redistributing the allocated keys accordingly) as opposed to halving the set of allocated keys directly. 
To illustrate this point, consider the case where there are hundreds of duplicate copies of $\kappa$. 
Then halving the key space means that all the hundreds of entries will move to one of the brand new data nodes (this is the unbalanced scenario we mentioned before). 
When a new data node is created, ALEX sequentially re-inserts the corresponding keys one by one to this new node.
Unfortunately, this sequence of operations follows the model-based insertion which means that all hundreds of repeated $\kappa$ entries will re-introduce conditions (1) $\&$ (2) in the brand new data node.  
The last piece of the puzzle is the following observation: if the conditions (1) and (2) in the new data node are severe enough to increase the cost and result in yet another catastrophic event, the newly introduced data node will have a high cost that will trigger another split. 
If ALEX reaches the above state by a single insertion, then the host will experience a recursive sequence of splits where each of them doubles its parent's pointer array. 
This cascading effect of an exponentially growing memory allocations will only stop when the host machine runs out of memory. 

\noindent\textbf{An illustration of the Attack. }
Suppose we have a data node responsible for the key space $[0,20)$. 
Let's assume, for the sake of the illustration, that the parent node has a pointer array of length two and that the key $\kappa=13$ has 200 copies that appear in the right data node (see Figure~\ref{fig:ce}(a)). 
Then, when the attacker inserts key $\kappa=13$ again, the cost that the cost model calculates increases so much that it triggers a catastrophic event and, consequently, a sideways split. 
As a result, the old light-red colored data node that was responsible for key space $[10,20)$ is split to the left (red colored again) data node responsible for key space $[10,15)$ and a right data node responsible for $[15,20)$. 
Given that the parent node only had a single pointer to the old data node, the sideways split triggers a doubling of the parent's pointer array (see Figure~\ref{fig:ce}(b)). 
The crucial part is that all 201 entries of key $13$ moved to the new left data node in Figure~\ref{fig:ce}(b); therefore, the cost of this new data is too high, which triggers yet another split (see Figure~\ref{fig:ce}(c)). 
This time the key space of the data node with the 201 entries of $13$ has key space $[12.5,15)$, and due to the lack of available pointers on the parent node, another doubling of the pointer array takes place. 
Figure~\ref{fig:ce}(d) illustrates that the pointer array (colored in dark red) grows exponentially, and all entries of $13$ move to a data node with a refined key space partition. 
Overall, the ``batch'' of copies of $\kappa=13$ move to data nodes with more and more fine-grained key space: 
\vspace{-0.08cm}
\begin{equation*}
\vspace{-6pt}
\resizebox{.47\textwidth}{!}{%
$
\begin{aligned}
[10,20)&\rightarrow [10,15) \rightarrow [12.5,15)\rightarrow [12.5,13.75) \rightarrow [12.5,13.125) \rightarrow [12.8125,13.125) \\ &\rightarrow [12.96875,13.125)\rightarrow [12.96875,13.046875)\rightarrow [12.96875,13.0078125)\rightarrow \ldots
\end{aligned}$
}
\label{eq:mck}
\end{equation*}

\vspace{-2pt}
\subsubsection{Attack Method} 
\label{internal_node_am}
For the internal node attack, we consider a black-box setup where the attacker has no knowledge of the key distribution of the targeted ALEX structure.
To mount the attack, the adversary generates a key that falls anywhere in the key range and inserts the duplicate key to ALEX. 
Typically, a few hundred duplicate insertions are sufficient o trigger OOM on the host machine. 
Detecting this attack can be challenging since the attacker can interleave duplicate insertions with legitimate workload traffic, leading to an eventual OOM crash of the ALEX database server. 

\vspace{-2pt}
\subsubsection{Comparison with Concurrent Work}
\label{sec:internal_node_comparison} 
In the following, we highlight the differences between our proposed attack and that from a recently released concurrent manuscript by Schuster~et~al.~\cite{lss_arxiv22} (referred to as {\small\texttt{SZEGP}}). 
The effectiveness of our approach, compared to {\small\texttt{SZEGP}}, is validated through our experiments in \cref{sec:internal_node_eval}, where our attack achieves comparable damage with up to $300,000\times$ fewer keys. 
{\small\texttt{SZEGP}} is parameterized by $N,K$ and works as follows: 
(1)~being a black-box attack, {\small\texttt{SZEGP}}  randomly samples $N$ keys from the original dataset to roughly estimate the global key distribution; 
(2)~{\small\texttt{SZEGP}} picks a random position in the key range and inserts $K$ \emph{consecutive} keys left to the position. If the keys are floating point, then {\small\texttt{SZEGP}} generates keys in $10^{-13}$ increments, if the keys are integers then it generates keys in $1$ increments. 
(3)~{\small\texttt{SZEGP}} repeats Step\#2 until the budget is exhausted or an OOM error is triggered. 

\noindent\noindent\textbf{Comparison with {\small\texttt{SZEGP}}.} 
While our internal-node space ACA and {\small\texttt{SZEGP}} share the common goal of doubling the size of parent nodes' pointer array, our attack uses fundamentally different insights to deliver a much more effective attack, as outlined below:
\begin{itemize}[noitemsep,leftmargin=*]
\item\textbf{Identifying a Data Node}: Our attack can be applied to \emph{any} data node, whereas {\small\texttt{SZEGP}} is only effective when applied to the \emph{left-most} data node of a sub-tree. 
Thus, {\small\texttt{SZEGP}} needs to guess a key-area for the cluster and blindly (due to the black-box setting) hit a range of the key distribution  with a very specific ALEX sub-structure.  
More often than not, {\small\texttt{SZEGP}} does not meet the above stringent requirement and as a result ``wastes'' valuable budget.
\item{\bf Effect on Landed Data Node}: Our attack guarantees that \emph{all} inserted keys will end up in the same data node (due to the fact that all keys are identical), which translates to doubling the same pointer array, and thus, delivering a faster OOM. 
Whereas the consecutive keys inserted by {\small\texttt{SZEGP}} may \emph{only partially} land to the left-most data node while the rest spill to a nearby data node. 
\item{\bf ALEX's Handling of Insertions}:  Our attack guarantees that \emph{all} inserted keys will be moved to a single new data node after the split, whereas  {\small\texttt{SZEGP}} (consecutive) inserted keys may be redistributed to both new data nodes from the split. 
Consequently, for {\small\texttt{SZEGP}}, ALEX may indeed reduce the cost by a sideways split.
\item{\bf Requirement on Adversarial Budget}: Our attack operates using a small adversarial budget; typically, a few hundred is sufficient to trigger an OOM event. 
Conversely, {\small\texttt{SZEGP}} relies on adversarial insertions falling into the leftmost data node within a sub tree; to accomplish such an allocation their attacker needs to use a significantly larger budget.
\end{itemize}

\vspace{-2pt}
\subsubsection{Evaluation} 
\label{sec:internal_node_eval}

We evaluate the internal node attacks on AWS EC2 {\small\texttt{m5}} virtual machine (VM) instances with the RAM configuration ranging from 64GB to 384GB. 
We use the four datasets from the previous subsection: {\small\texttt{longitudes}}, {\small\texttt{longlat}}, {\small\texttt{YCSB}}, and {\small\texttt{lognormal}}.

To compare, we initialized an ALEX instance with 10~million legitimate keys from the dataset using {\small\texttt{bulk\_load()}}, and then mounted the attack. 
For {\small\texttt{SZEGP}}, we used the same parameter configurations as the ones used in \cite{lss_arxiv22}, where $N = 1,000$ and $K = 10,000$.
We recorded the number of adversarial insertions used and the total amount of memory consumed by ALEX under both types of attacks. 


\begin{table}[t]
    \centering
    \caption{Effectiveness of the space ACA on internal node.  
    \textit{\textmd{In the {\small\texttt{OOM}} rows, {\cmark} denotes a triggered OOM. Otherwise, the values show the memory consumption after the specified budget is used.}}}
    \vspace{-10pt}
    \label{tbl:internal_aca}
    \resizebox{\columnwidth}{!}{%
                   \begin{tabular}{ll r cccc cccc}
            \multicolumn{2}{c}{} & \multicolumn{1}{c}{} & \multicolumn{4}{c}{\textbf{Our attack}} & \multicolumn{4}{c}{\textbf{Schuster et al.} ({\small\texttt{SZEGP}})}\\
            \cmidrule(lr){4-7} \cmidrule(lr){8-11}
            
            \multicolumn{2}{c}{} & \multicolumn{1}{r}{\textbf{Host memory (GB)}} & \textbf{64} & \textbf{128} & \textbf{256} & \textbf{384} & \textbf{64} & \textbf{128} & \textbf{256} & \textbf{384} \\

            \Xhline{2\arrayrulewidth}
            
            \multicolumn{2}{l}{\multirow{2}{*}{\textbf{{\texttt{Longitude}}}}}
                        & \textbf{OOM}  & \cmark & \cmark  & \cmark  & \cmark  & \cmark & \cmark  & 244GB  & 244GB \\
                        & & \textbf{Budget} & 423 & 479  & 472  & 441  & 63M & 116M & 200M  & 200M  \\
                        
            \hline
                        
            \multicolumn{2}{l}{\multirow{2}{*}{\textbf{{\texttt{Longlat}}}}}
                        & \textbf{OOM}  & \cmark & \cmark  & \cmark  & \cmark  & \cmark & \cmark  & \cmark  & \cmark \\
                        & & \textbf{Budget} & 428 & 441  & 437  & 437  & 2069 & 2069 & 2069  & 2069  \\
                        
            \hline
            \multicolumn{2}{l}{\multirow{2}{*}{\textbf{{\texttt{YCSB}}}}}
                        & \textbf{OOM}  & \cmark & \cmark  & \cmark  & \cmark  & \cmark & \cmark  & \cmark  & \cmark \\
                        & & \textbf{Budget} & 662 & 653  & 692  & 679  & 1024 & 1024 & 1024  & 1024  \\
                        
            \hline
            \multicolumn{2}{l}{\multirow{2}{*}{\textbf{{\texttt{Lognormal}}}}}
                        & \textbf{OOM}  & \cmark & \cmark  & \cmark  & \cmark & 5GB & 5GB  & 5GB  & 5GB \\
                        & & \textbf{Budget} & 526 & 589  & 564  & 551  & 200M & 200M & 200M  & 200M  \\
                        
            \end{tabular}
    } 
    \label{table:internal_aca}
\end{table}

\begin{figure}[t]
\begin{center}
    \subfigure[Memory consumption vs. insertions.] {
    \includegraphics[width=.212\textwidth]{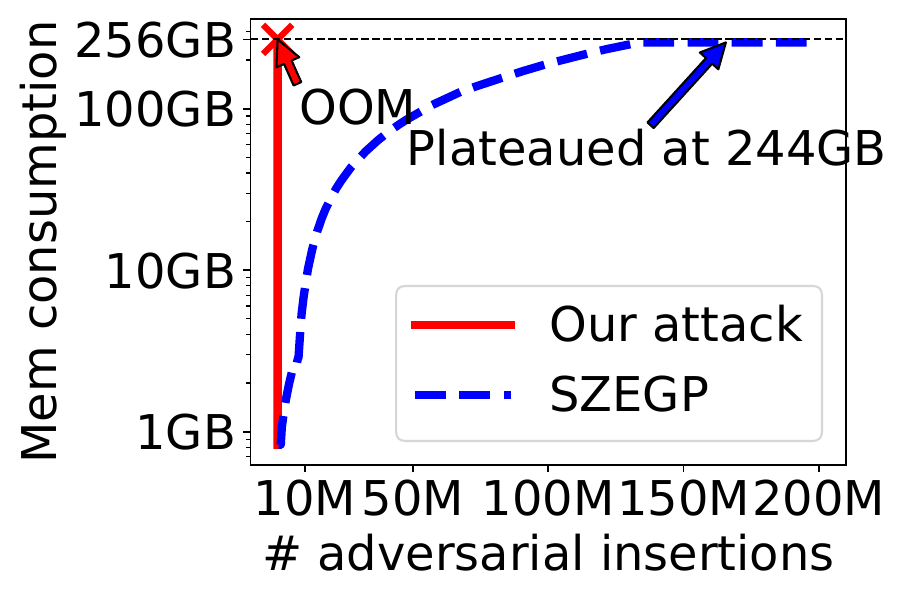}
    \label{fig:internal_km}
    }
    \hspace{-5pt}
    \subfigure[Memory consumption vs. runtime.] {
    \includegraphics[width=.212\textwidth]{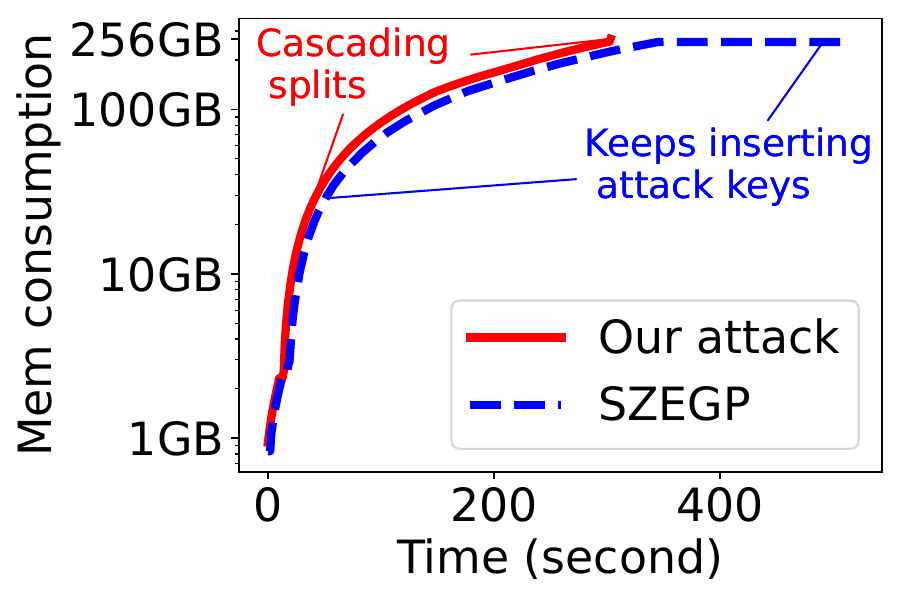}
    \label{fig:internal_tm}
    }
    \vspace{-10pt}
    \caption{Memory usage as a function of the number of adversarial insertions for the {\small\texttt{longitudes}} dataset.}
    \label{fig:longitudes_trend} 
\end{center}
\end{figure}

The results presented in Table~\ref{tbl:internal_aca} comprehensively compare our method and  {\small\texttt{SZEGP}}. 
Our attack method consistently induced an OOM event on all VM configurations tested across all four datasets, with only 423-692 adversarial insertions. 
On the other hand, {\small\texttt{SZEGP}} triggered OOM in only two out of the four datasets, i.e.,  {\small\texttt{Longlat}} and {\small\texttt{YCSB}}, by using $3.83\times$ and $0.55\times$ more budget than our method to exhaust 64GB of host memory. 
The difference in the effectiveness between the two attack methods is striking in datasets {\small\texttt{Lognormal}} and {\small\texttt{Longitudes}}, where {\small\texttt{SZEGP}} requires more than $10^5\times$ budget and in most cases, it doesn't trigger an OOM. 
In Figure~\ref{fig:longitudes_trend}, we take a more detailed view of the comparison between the two methods on {\small\texttt{Longtitudes}}. 
Figure~\ref{fig:longitudes_trend} shows the memory consumption as a function of the number of adversarial insertions and runtime on a 256GB EC2 VM. The proposed attack brings the data node to a state where \emph{a single insertion} will exhaust all available memory regardless of whether the host memory is 1GB or 1TB. 
Figure~\ref{fig:internal_km} illustrates this point: no matter how much the defender increases the host's memory, once the trigger insertion takes place all memory is consumed.
Our space ACA attack only required 472 insertions of duplicate keys to trigger cascading splits at a memory consumption rate of $0.84$~GB/sec until all the 256GB RAM is exhausted  (Figure~\ref{fig:internal_tm}).  
{\small\texttt{SZEGP}} required contiguous and cumulative adversarial insertions, ultimately reaching a memory usage plateau of around 244GB with $3\cdot10^{5}\times$ more budget.
For the {\small\texttt{Lognormal}} dataset, {\small\texttt{SZEGP}} used up all the 200~million budget keys but only caused about 5GB memory increase. 
These results demonstrate the effectiveness of our method. 

\noindent\noindent\textbf{Potential Mitigation.} One may argue that this attack can be mitigated by using a capacity-based partitioning mechanism that produces equally sized data nodes. This defense mechanism could potentially reduce the cost of a data node, as the duplicate keys will eventually be spread across multiple data nodes. However, incorporating capacity-based partitioning will break ALEX's sophisticated design for the following reasons:
(1)~the ultimate goal of key space partitioning is to make sure that each data node covers a sub-key space whose key distribution is roughly linear;
and (2)~key space partitioning enables ALEX to use simple linear models for internal nodes. 
Capacity-based partitioning would have required internal nodes to use expressive, non-linear models as it is highly likely that the key distribution within a child data node is non-linear. 
Another possible defense is by disabling the option for duplicate keys. However, this will inevitably hurt the versatility and flexibility of ALEX as a core database index structure. 
In the next section we discover the feasibility of mitigating our attack by changing how ALEX handles catastrophic events.

\begin{figure*}[t]
\begin{center}

\subfigure[{\small\texttt{YCSB}}.] {
\includegraphics[width=.22\textwidth]{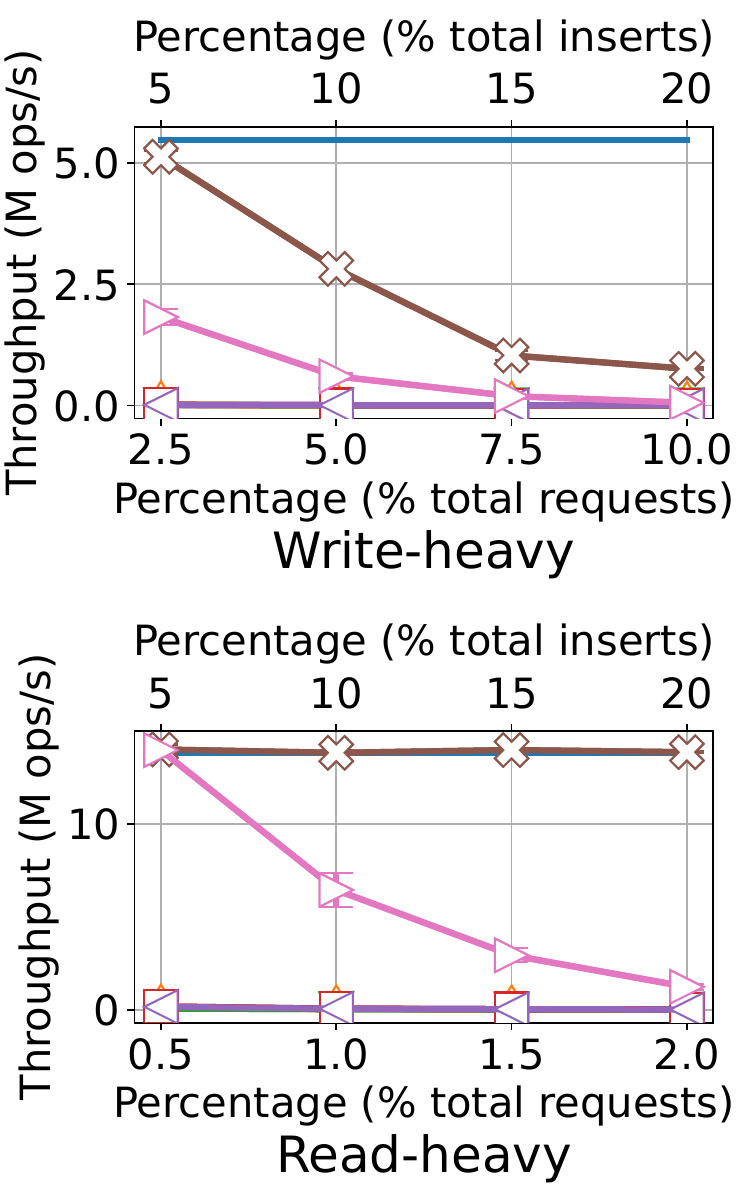}
\label{fig:ac_time_ycsb}
}
\hspace{-8pt}
\subfigure[{\small\texttt{Lognormal}}.] {
\includegraphics[width=.22\textwidth]{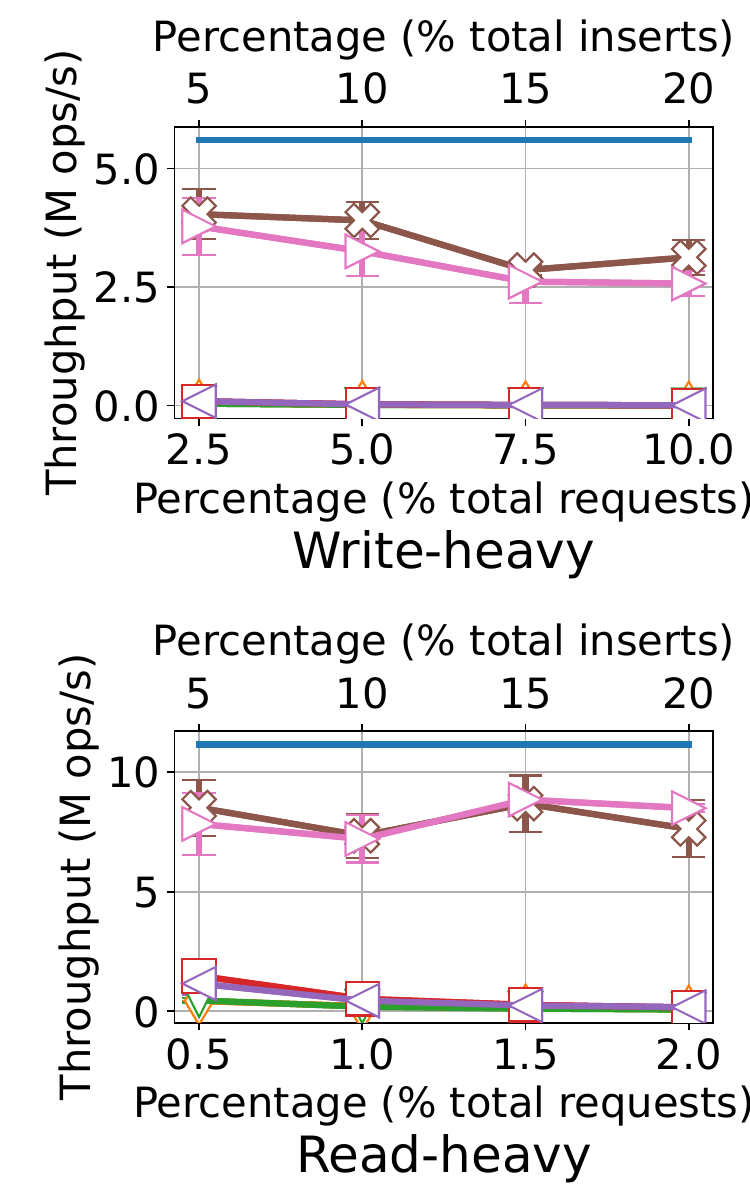}
\label{fig:ac_time_lognormal}
}
\hspace{-8pt}
\subfigure[{\small\texttt{Longitude}}.] {
\includegraphics[width=.22\textwidth]{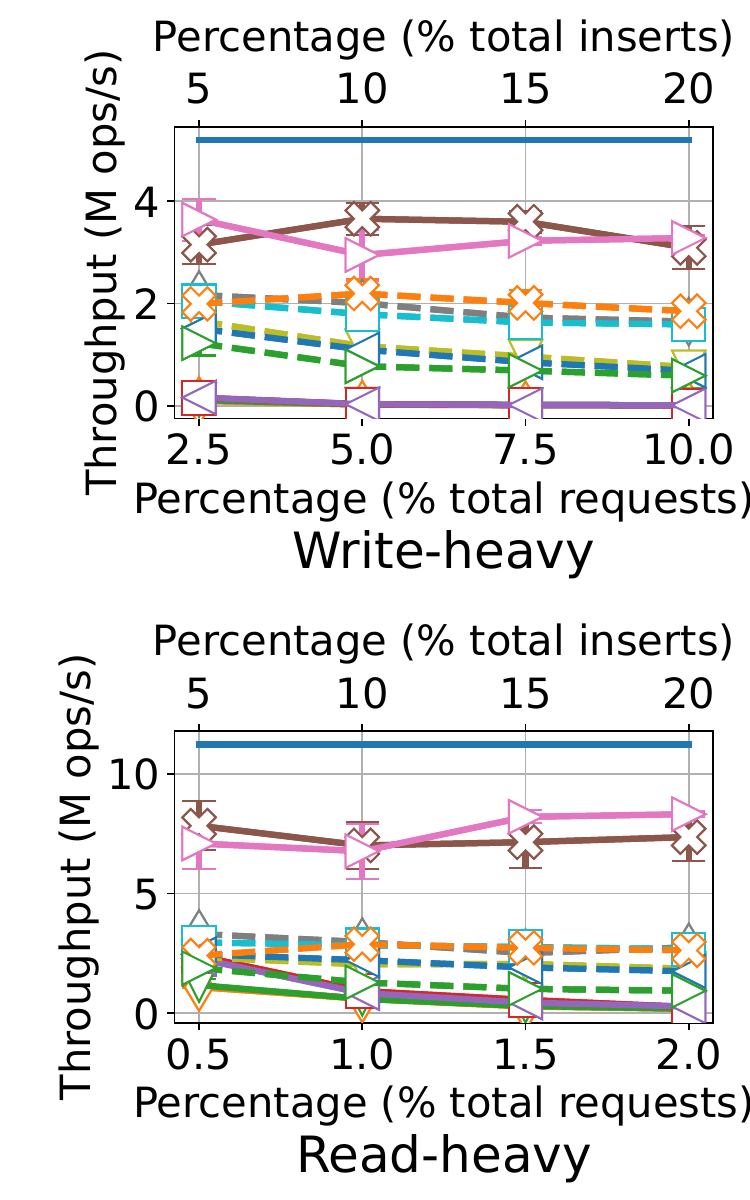}
\label{fig:ac_time_longitude}
}
\hspace{-8pt}
\subfigure[{\small\texttt{Longlat}}.] {
\includegraphics[width=.31\textwidth]{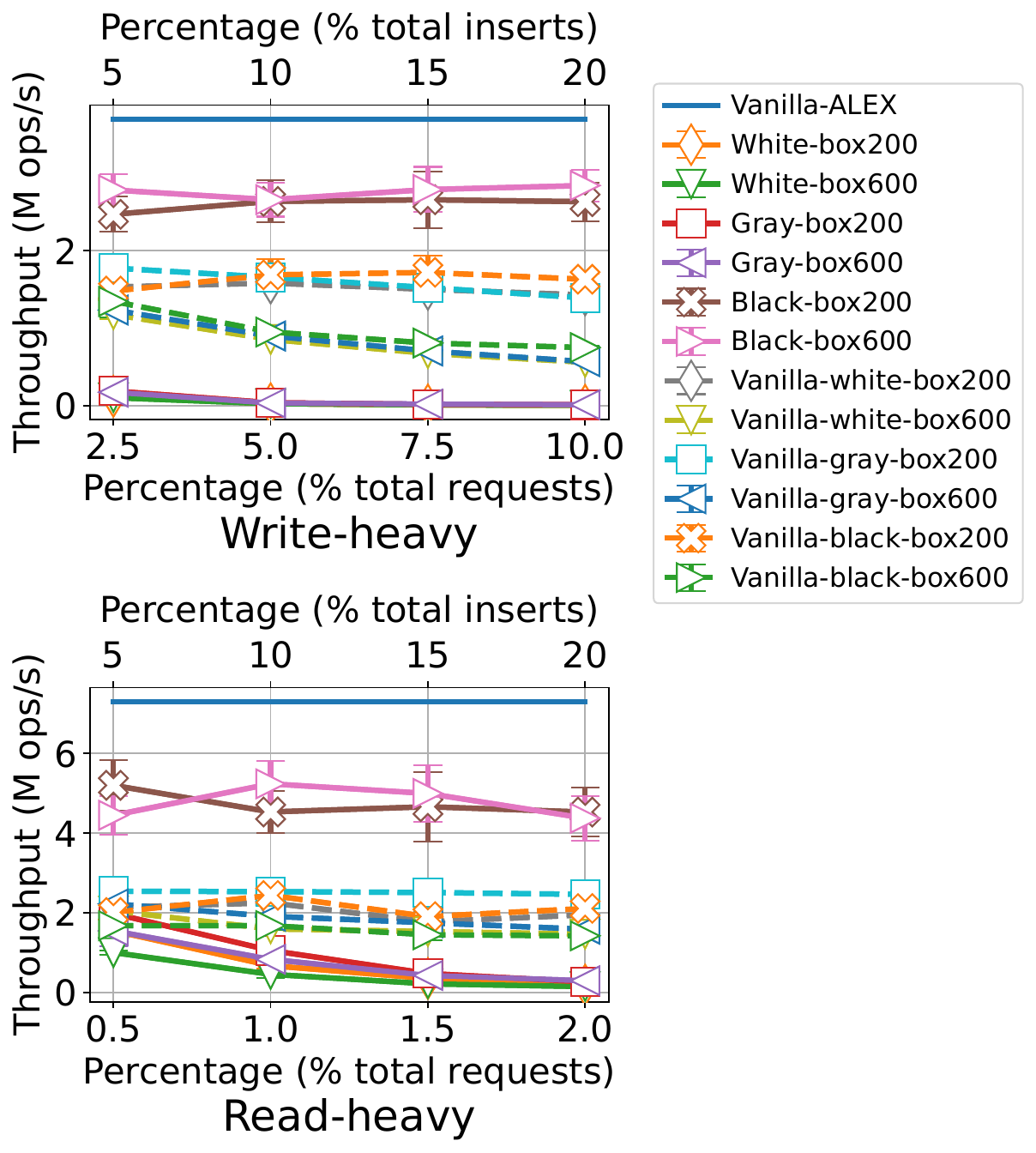}
\label{fig:ac_time_longlat}
}

\vspace{-12pt}
\caption{
The throughput for different attack settings and percentages of adversarial insertions for vanilla ALEX and modified ALEX.
\textit{\textmd{
We varied the percentage of adversarial insertions with respect to (w.r.t.) the total number of inserts from $5\%$ to $20\%$.
The top $X$-axis shows the ratios of adversarial insertions to the total number of insertion requests in the workload.
The bottom $X$-axis shows the normalized ratios of adversarial insertions to the total number of requests in the workloads: e.g., a ratio of $2.5\%$ for a write-heavy workload w.r.t. the bottom $X$-axis means, out of the $10M$ requests, $2.5\%$ of them are adversarial insertions generated by the attacker; by referring to the top $X$-axis, $2.5\%$ of total requests means $5\%$ of total inserts as our write-heavy workload has a $50\%:50\%$ read:write ratio.
{\small\texttt{Vanilla-ALEX}} is our baseline performance without any adversarial manipulation.
To measure the performance of our attacks, we applied them to both vanilla/original ALEX and modified ALEX.
{\small\texttt{Vanilla-white-box200}} denotes a white-box attack applied to the vanilla ALEX. 
{\small\texttt{White-box200}} means a white-box attack mounted against the modified ALEX.
The values after the attack setting denote the batch sizes; e.g., {\small\texttt{White-box200}} means the white-box attack with a batch size of 200. Each data point is the average of five runs with error bars showing the min-max variance. 
}}}
\label{fig:ac_time_evals}
\end{center}
\end{figure*}

\vspace{-2pt}
\subsection{Time ACA}
\label{sec:time_aca}

In this subsection, we introduce a new category of ACA that aims to deteriorate ALEX's time performance. 
In~\cref{sec:internal_node_aca}, we introduced a space ACA that exploits the fact that catastrophic events trigger sideways splits. 
A potential mitigation of our space ACA is the following: in case of a catastrophic event, perform an expansion operation instead of a sideways split.  
As we show in the following, when expansions happen in the context of a catastrophic event, ALEX needs to retrain the linear regression model of the corresponding data node. 
However, retraining is a time-consuming operation (a fact that we capitalize on in our time ACA). 
In this subsection, we analyze a \emph{hypothetical} scenario where ALEX is patched to mitigate our space ACA vulnerability and now performs an expansion instead of a sideways split in case of a catastrophic event\footnote{We also tested the vulnerability of vanilla ALEX to our time ACA in \cref{aca_time_eval}.}. 

\vspace{-2pt}
\subsubsection{Exploitable Design Choice} 
\label{sec:temporal_aca_vul}
Recall that a catastrophic event happens when ALEX detects that the cost of a data node is too high. 
A reason that contributes to the high cost is that  the linear regression model prediction is no longer accurate to support efficient lookups and insertions.
While applying an expansion instead of a split thwarts our attack from~\cref{sec:internal_node_aca}, it will cause inaccurate models that result in significant runtime performance cost due to increased training cost.
An attacker can exploit this observation of the modified ALEX by strategically inserting sequences of densely clustered  
keys to carefully chosen data nodes to cause a discrepancy between the key distributions and the linear models of the data nodes. 

\vspace{-2pt}
\subsubsection{Attack Method} 
\label{sec:temporal_aca_am}
The goal of an attacker is to leverage the above-described observation to deteriorate the performance of ALEX by increasing its time performance~\cite{acatcp1_usenixsec05, acatcp2_linux18, aca_usenixsec03}. 
This will result in:
(1)~query latency increase when serving highly-concurrent requests,
and (2)~ultimately, a DoS when the request processing capability offered by the database tier is not able to catch up with the speed of query backlog accumulation. 
Data node expansion implies that the current model of the data node is no longer accurate, thus a model retraining is required.
However, retraining will not fix the inaccuracy issue, simply because the key distribution of the data node \emph{is no longer linear}. Such inaccuracy will cause more exponential searches for lookups and more memory shifts for insertions. 
More importantly, expanding a data node in ALEX is a time-consuming process due to the overhead of linear model retraining and memory copy. 
Therefore, an effective adversarial strategy for degrading ALEX's performance is to continuously trigger ``catastrophic'' expansions (see \cref{sec:prelim_alex}) in large data nodes. 
To this end, we investigate time ACA strategies in three different settings: white-box, gray-box, and black-box. 

The white-box setting assumes that the attacker has complete knowledge of the target ALEX, including keys and the global key distribution. In this case, the attacker 
targets the largest data node. Specifically, the attacker inserts
a total number of $N$ attacking keys into the middle position of the largest segment within the largest data node.
The attacker spreads the budget of $N$ keys across batches with a size of $B$, where $B$ is set to 200 and 600 in our tests and all keys within a single batch are consecutive keys with an incremental difference of $10^{-13}$ for {\small\texttt{double}}-typed keys and $1$ for {\small\texttt{int}}-typed keys.

In the gray-box setting, the attacker knows the approximate distribution of the keys but does not know the exact keys. With the distribution, the attacker utilizes a kernel density estimation (KDE)~\cite{tool_sklearn} method to construct a substitute ALEX structure that approximates the structure of the target (original) ALEX. The substitute ALEX is designed to have the same size as the target ALEX and contains different keys, which fall within the same key range and follow the same key distribution. 
Subsequently, the attacker devises an attack plan by generating attacking keys targeting the largest data node in the substitute ALEX, same as the strategy used in the white-box attack. 
The attacker then inserts the $N$ budget keys to the target ALEX in $N/B$ batches during the workload.  

Lastly, the black-box setting assumes that the attacker does not know about the key distribution but can probe the existing key range of ALEX via sampling. The attacker then randomly choose $N/B$ locations from within the estimated key range and insert the $N/B$ batches of attacking keys to mount the time ACA. 
\vspace{-2pt}
\subsubsection{Evaluation}
\label{aca_time_eval}

We evaluated the time ACAs on an EC2 {\small\texttt{m5}} VM instance with 96~vCPUs and 384GB memory. We used the average throughput as the performance metric.
We mounted our time ACA against both the vanilla ALEX (the original ALEX) and our modified ALEX (patched to mitigate the space ACA vulnerabilities as discussed at the beginning of \cref{sec:time_aca}). We conducted the experiments using a write-heavy and a read-heavy workload drawn from the four datasets.
The write-heavy workload consists of $50\%$ inserts and $50\%$ lookups, while the read-heavy workload has an insert:lookup ratio of $10\%:90\%$. Both workloads have  a total of 10~million operations. The lookup keys were selected from all existing keys in one of the four datasets, following a Zipfian distribution. 
Given a dataset, we first initialized an ALEX by using the first 10~million keys via {\small\texttt{bulk\_load()}} and then initiated the 10-million-request workload. We issued a total of 10 batches for all attack settings.

Figure~\ref{fig:ac_time_evals} shows the throughput comparisons between a modified ALEX (with expansion-only mode as described in \cref{sec:time_aca}) and the vanilla ALEX in different attack settings with baseline.
For write-heavy workloads, the white-box attacks and gray-box attacks in modified ALEX achieved an average throughput degradation between $31\times$ and $1,641\times$ across all batch size and budget parameters, compared to the baseline ALEX.

\begin{figure}[h]
\begin{center}
    \subfigure[Read-heavy Workload.] {
    \includegraphics[width=.215\textwidth]{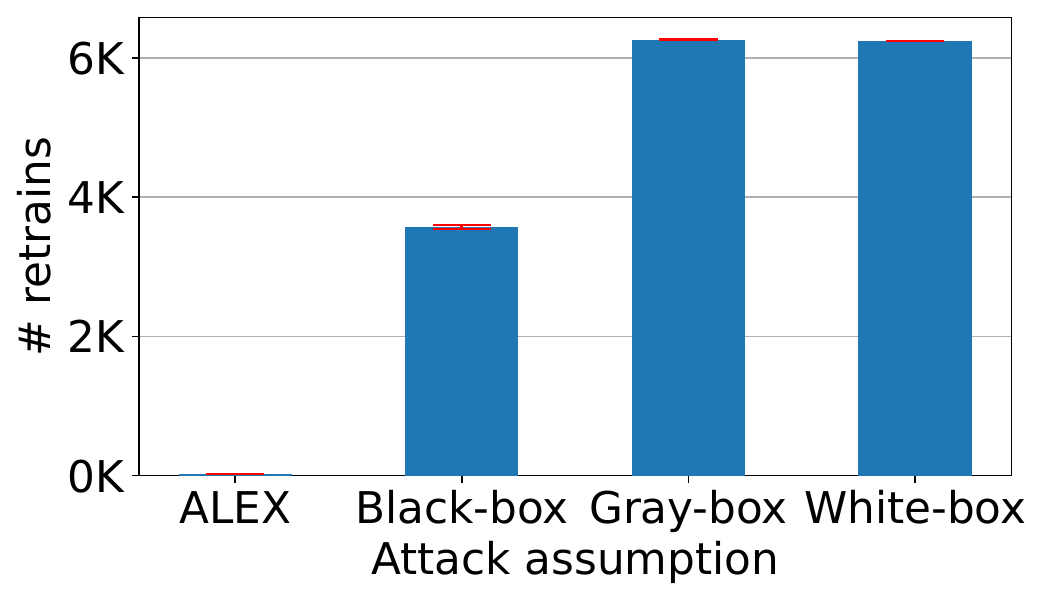}
    }
    \hspace{-4pt}
    \subfigure[Write-heavy Workload.] {
    \includegraphics[width=.215\textwidth]{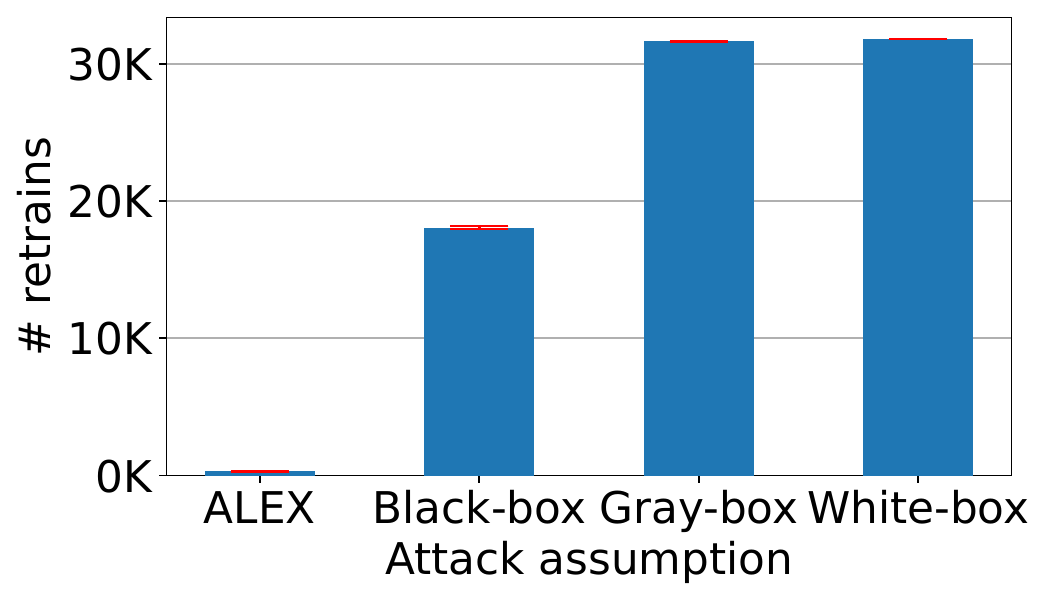}
    }
    \vspace{-10pt}
    \caption{Number of retrains under various attack assumptions for {\small\texttt{Longitudes}}. Adversarial insertions are $20\%$ of the total number of insertions.  
    \textit{\textmd{Each bar in the plot represents the average of five runs with red-colored error bars showing the variance.}}} 
    \label{fig:longitudes_num_retrains}
\end{center}
 \vspace{-5pt}
\end{figure}

This is because all attack keys were inserted into the largest data node, causing the biggest damage to performance. 
Black-box attacks, on the other hand, saw relatively fluctuating performance trends for {\small\texttt{longitude}}, {\small\texttt{longlat}}, and {\small\texttt{lognormal}} (Figure~\ref{fig:ac_time_longitude}--\ref{fig:ac_time_lognormal}), due to the fact that the ALEX trees constructed under these three datasets had many data nodes. As a result, a single batch of adversarial insertions may end up falling into more than one data node, with reduced likelihood of triggering a ``catastrophic'' expansion.  
Interestingly, the performance degradation on {\small\texttt{YCSB}} increases as both the batch size as well as the attack budget increase, as shown in Figure~\ref{fig:ac_time_ycsb}, because 
there were much fewer data nodes used in ALEX to store the linearly distributed keys in the {\small\texttt{YCSB}} dataset, which increased the chances of a ``catastrophic'' expansion being triggered and minimized the attack budget wastage. As a result, the throughput of {\small\texttt{Black-box600}} decreases sharply as the budget ratio increases; with only $2\%$ of requests being adversarial insertions, our black-box-based time ACA caused a performance degradation of up to $11\times$
compared to the baseline case for the read-heavy workload (see the bottom sub-figure of Figure~\ref{fig:ac_time_ycsb}), almost approaching the same level of degradation caused by a white-box attack.
For attacks against vanilla ALEX, our time ACA achieved $2.1\times$ to $8.8\times$ performance degradation on write-heavy workload and up to $11.99\times$ on read-heavy workload for {\small\texttt{longitude}} compare to baseline ALEX (note that we were not able to collect the throughput results as our time ACA triggered an OOM event on {\small\texttt{YCSB}} and {\small\texttt{lognormal}} against the vanilla ALEX).
Figure~\ref{fig:longitudes_num_retrains}
provides additional insight into the throughput performance drops under different attack assumptions.
Both gray-box and white-box approaches incur significant higher
number of retrains when compared to ALEX with legitimate workload and the black-box setting.
This indicates that the decrease in throughput can be primarily attributed to the number of retrains.

\vspace{-3pt}

\noindent\textbf{Potential Mitigation. }One possible defense is to apply more robust ML models (e.g., polynomial regression) to represent the empirical CDF of keys at data-node level. 
An important caveat of this mitigation strategy is that applying more expressive ML models deviates from the original motivation of using piece-wise linear regression models to approximate empirical data distribution CDF, and the time overhead of model training and inference might outweigh the robustness these complicated models introduce. 
\vspace{-2pt}
\section{Related Work}
\label{sec:related_work}

\noindent\textbf{Other Learned Index Structures.}
LIS use ML models to replace traditional index structures based on the observation that index structures can be viewed as a CDF~\cite{lis_sigmod18}.
FITING-Tree~\cite{ftree_icmd19} employs piece-wise linear functions with a predetermined error bound during construction. In terms of dynamism of LIS, a method of minimizing error caused by updating index was proposed by Hadian and Heinis in~\cite{1st_aidm19}.
The work by Tang~et~al.~\cite{tanglearned_arvix19} studies the distribution of the workload and propose to re-train the model as the queries pattern change.
Other works propose learned multi-dimensional index structure~\cite{tsunami_vldb20, mdlis_sigmod20}.
PGM~\cite{pgm_vldb20} proposes models that utilize an auto-tuned RMI design to optimize space and latency while supporting updates by using an index-level shared write buffer (a  strategy similar to the LSM-tree structure~\cite{lsm_tree}).
Based on PGM, RadixSpline~\cite{radixspline_aidm20} proposes an RMI that features an alternate linear interpolation based indexing.
Lately, a line of works aim to improve dynamic learned index structure in robustness, concurrency, persistence, and the capability to operate under limited DRAM resources~\cite{apex_vldb21, finedex_vldb21, lipp_vldb21, film_vldb22, plin_vldb22, nfl_vldb22}.
Inspired by LIS, \textsc{SageDB}~\cite{segadb_cidr19} is a database system that adapts to an application through code synthesis and learning techniques. 
A set of works benchmark the performance of learned index structures~\cite{sosd_arxiv19, bli_vldb20, lis_benchmarking_vldb22, lis_benchmarking_vldb23}.
Our work fills a missing gap in the literature by investigating ACAs against emerging dynamic learned index structure~\cite{alex_sigmod20, apex_vldb21}.

\noindent\textbf{Algorithmic Complexity Attacks.}
ACAs were initially introduced~\cite{aca_usenixsec03} to exploit worst-case algorithmic design choices in common data structures such as hash tables.
Several other studies explored ACAs across diverse applications, such as hash tables~\cite{acaht2_springer07, acaht1_tc13}, regular expression matching~\cite{acarm_idtacas17, rescue_ase18, acarm_esecfse18}, automata-based multi-string pattern matching~\cite{acasm_springer12}, PDF (portable document file)  decompression~\cite{acadpf_blackhat}, and TCP reassembly~\cite{acatcp1_usenixsec05, acatcp2_linux18}. 
\cite{perffuzz_issta18, slowfuzz_ccs17} propose solutions that automatically detect inputs that cause algorithmic complexity vulnerabilities. 
On the defense side, \textsc{SurgeProtector}~\cite{surgeprotector_sigcomm22} 
proposes a general framework to make network functions resilient against ACAs. None of these works explore vulnerabilities of dynamic learned indexes, which we do in this work. 

\noindent\textbf{Data Poisoning Attacks.}
The literature on data poisoning attacks~\cite{clustering_adversarial_aisec13, feature_poisoning_icml15, yang2017generative, poisoning_byzantine_sec20}
focuses on adversaries who intentionally augment the training data to manipulate the outcomes of predictive models. For instance, Biggio~et~al.~\cite{clustering_adversarial_aisec13} introduced maliciously crafted training data to alter the decision function of support vector machines (SVMs) and increase the test error. Yang~et~al.~\cite{yang2017generative} proposed gradient-based methods to generate poisoning points for neural networks. Suciu~et~al.~\cite{gtepa_usenixsecurity18} presented a framework to evaluate realistic adversaries conducting poisoning attacks on machine learning algorithms. 
The study by Jagielski~et~al.~\cite{mml_arxiv21} 
proposes an optimization framework for poisoning attacks on linear regression and introduced a defense mechanism named Trim. 
The ACAs we propose in this work are different from data poisoning attacks. 
The objective of poisoning attacks is to maximize the mathematical function that captures the error of the ML model, whereas our ACAs maximize the use of critical resources based on how the system is designed (which cannot always be captured by a simple error function).

\noindent\textbf{Comparison with {\small\texttt{KRT}}}. 
The poisoning attacks presented in~\cite{poisoning_lis_sigmod22} focus on \emph{static} learned index structures. 
This means that all inserted keys in~\cite{poisoning_lis_sigmod22} are chosen during the initialization. 
On the contrary, in this work our attacks are performed on a \emph{dynamic} LIS and the keys are chosen based on the current state of the LIS. 
Another difference is that at the core of the attacks~\cite{poisoning_lis_sigmod22} is the observation that injections of keys in cumulative distribution functions cause a cascading error effect. 
The above error cannot be dealt with at runtime due to the static nature of the LIS.
On the contrary, the attacks in this work are centered around the runtime expansion mechanisms when dealing with full-occupancy of a node or increased errors.  
These mechanisms only appear in the dynamic setting.

\vspace{-2pt}
\section{Conclusion}
\label{sec:conclusion}

We present a comprehensive study that focuses on the security aspects of an emerging dynamic learned index structure. We propose new ACAs that exploit design choices that balance the trade-off in ALEX's memory usage and runtime efficiency. 
Our attacks aim to overload the memory and CPU resources. 
We evaluate the effectiveness of these ACAs through extensive experiments.
Our findings have demonstrated that our space ACAs cause out-of-memory with only a few hundred  adversarial insertions and our time ACAs lead to a significant degradation in throughput by up to three orders of magnitude, compared to ALEX.
Our findings can be used to inform the future generations of robust learned-index-based systems that are not prone to adversarial manipulations. 
\vspace{-3pt}
\begin{acks}
\label{sec:acknowledgements}
\end{acks}

We are grateful to the anonymous reviewers for their valuable feedback and comments. This work was sponsored in part by NSF grants: CCF-2318628, CCF-1919113, OAC-2106446, CNS-2154732, and the Meta Security Research Award.

\clearpage
\balance
\bibliographystyle{ACM-Reference-Format}

\bibliography{ref}
\end{document}